\documentclass[prd,aps,superscriptaddress,10pt,nobibnotes,notitlepage,twocolumn,longbibliography,nofootinbib,preprintnumbers]{revtex4-1}
\usepackage{wrapfig}
\usepackage[normalem]{ulem}
\usepackage{sidecap}
\usepackage[latin1]{inputenc}
\usepackage{graphicx}
\usepackage{float}
\usepackage{latexsym}
\usepackage{graphicx}
\usepackage{amssymb}
\usepackage{amsmath,mathtools}
\usepackage{amsfonts}
\usepackage{tikz} 
\usepackage{ragged2e}
\usepackage{stackengine}
\usepackage{subcaption}
\usepackage{scalerel}
\PassOptionsToPackage{colorlinks=true,citecolor=blue,hyperfootnotes=false}{hyperref}

\usepackage{orcidlink}
\usepackage{lipsum}

\usepackage[colorlinks=true,citecolor=blue,hyperfootnotes=false]{hyperref}
\usepackage{dcolumn}
\usepackage{textcomp}
\usepackage{xfrac}
\usepackage{slashed}
\usepackage{multirow}

\def\be{\begin{equation}}
\def\ee{\end{equation}}
\def\figs/B{B}
\def\bea{\begin{eqnarray}}
\def\eea{\end{eqnarray}}
\def\bg{\begin{eqnarray}}
\def\nd{\end{eqnarray}}

\def\log{{\rm log}}
\def\ln{{\rm ln}}

\def\beq{\begin{equation}}
\def\eeq{\end{equation}}

\usepackage{graphicx}
\usepackage{dcolumn}
\usepackage{bm}
\usepackage{hyperref}
\usetikzlibrary{decorations.pathmorphing}
\tikzset{snake it/.style={decorate, decoration=snake}}

\begin{document}
\preprint{%
  \parbox[t]{4.5cm}{\raggedleft
    IFT-UAM/CSIC-26-139\\
    MIT-CTP/5980}}

\title{Revisiting Critical Higgs Inflation: \\
Electroweak Couplings, CMB Observables, and Perturbative Unitarity}

\author{Sarah R.~Geller \orcidlink{0000-0003-4126-4662}}
 \email{sarah.geller@yale.edu}

\affiliation{\ignorespaces
    Department of Astronomy, Yale University, Kline Tower, 266 Whitney Avenue, New Haven, CT 06511, USA
}

\author{Sol Gutierrez-Lara \orcidlink{0009-0004-6333-1744}}
 \email{sgutierrezlara@college.harvard.edu}
\affiliation{%
Department of Physics, Jefferson Laboratory, Harvard University, Cambridge, MA 02138, USA
}%

\author{David I.~Kaiser \orcidlink{0000-0002-5054-6744}}
 \email{dikaiser@mit.edu}
\affiliation{%
Department of Physics, Massachusetts Institute of Technology, Cambridge, MA 02139, USA
}%

\author{Thomas Steingasser \orcidlink{0000-0002-1726-2117}}
 \email{thomas.steingasser@uam.es}
\affiliation{%
Departamento de F\'{\i}sica Te\'orica, Universidad Aut\'onoma de Madrid, and IFT-UAM/CSIC, Cantoblanco, 28049, Madrid, Spain
}%

\date{\today}

\begin{abstract}
We revisit the scenario of critical Higgs inflation, in which inflation is driven by the Standard Model Higgs boson with parameters chosen close to a critical point, giving rise to an inflection point in its potential. After reviewing this model, we perform a systematic study of its viability centered around three criteria: (1) consistency with electroweak-scale measurements, (2) ability to reproduce the cosmic microwave background (CMB) observables, and (3) sensitivity of the model predictions to unknown UV physics, based on a breakdown of perturbative unitarity. We first show that compliance with electroweak-scale measurements is controlled by the parameters in the Higgs potential at inflationary scales, also updating the literature values. We then demonstrate how for any choice of potential parameters, agreement with the CMB observables can be achieved through the choice of the nonminimal coupling, in both metric and Palatini formulations of gravity. However, we also find that there exist \emph{no} regions of parameter space, in either the metric or Palatini formulations, in which compliance with all three physical criteria is possible. While this does \textit{not} exclude the possibility of the Higgs field driving inflation, it means that predictions for any observable are in principle sensitive to the details of the model's UV completion.

\end{abstract}

\maketitle

\section{Introduction}\label{sec:intro}

In recent decades, fundamental physics has relied on two remarkable frameworks: the $\Lambda$CDM model of cosmology, which governs the evolution of the Universe on large scales~\cite{Bull:2015stt,Perivolaropoulos:2021jda,PeeblesCosmicCentury}, and the Standard Model of particle physics (SM), completed with the discovery of the Higgs boson at the LHC in 2012~\cite{ATLAS:2012yve,CMS:2012qbp}, which treats interactions among elementary particles~\cite{Gaillard:1998ui,SchwartzBook}. Despite each theory's empirical successes, a challenge remains to unite them by identifying the microphysical ``ingredients'' that can account for such cosmological phenomena as the mechanism responsible for early-universe inflation. The simplest such mechanism is slow-roll inflation, in which inflation is driven by the potential energy of one or more scalar fields as they slowly evolve within their potential. (See Refs.~\cite{Guth:2005zr,Lyth:2009imm,Bassett:2005xm,Guth:2013sya,Martin:2013tda,Baumann:2022mni,Kinney:2026qtr} for reviews.)

It is well-established that slow-roll inflation can be achieved in agreement with all existing cosmological observations by a large number of models \cite{Martin:2013tda,Chowdhury:2019otk,Martin:2024qnn}. Many of these models predict additional observable but as yet unobserved features, such as a spectrum of primordial gravitational waves that may be within range of future detectors \cite{Allahverdi:2020bys,Domenech:2021ztg}, modest primordial non-Gaussianity \cite{Chen:2010xka,Renaux-Petel:2015bja,Celoria:2018euj}, nontrivial running of the spectral index of primordial curvature perturbations~\cite{Adshead:2010mc,Munoz:2016owz,Hardwick:2018zry,Easther:2021rdg,Bahr-Kalus:2022prj,Lorenzoni:2024krn}, and, in some instances, the production of primordial black holes \cite{Escriva:2022duf,Ozsoy:2023ryl}. Whereas such observables would provide valuable information about the dynamics of inflation, they would not suffice to identify the inflaton field itself. Whether successful inflation may be accomplished with ingredients that are already included within the established framework, or requires some new physics beyond the SM, remains an open question.

An intriguing and economical possibility is to identify the SM Higgs field as the inflaton. ``Higgs inflation''~\cite{Bezrukov:2007ep} holds particular appeal since the Higgs field is now known to exist, and its properties have been measured to high accuracy. (See Ref.~\cite{Rubio:2018ogq} for a review of Higgs inflation and its variants.) Models of Higgs inflation typically include a nonminimal coupling of the complex Higgs doublet ${\bf H}$ to the spacetime Ricci curvature scalar $R$ of the form $\xi {\bf H}^\dagger {\bf H} R$, with some dimensionless coupling constant $\xi$. Such couplings arise unavoidably through loop corrections for any self-interacting scalar field in a curved spacetime \cite{Birrell:1982ix,Buchbinder:1992rb,Parker:2009uva}. 

From a purely cosmological perspective, this scenario is remarkably successful. Higgs inflation is capable of explaining all CMB observables by choosing just one parameter, the nonminimal coupling $\xi$. In fact, in the limit $\xi \gg 1$, the original, tree-level Higgs inflation model behaves as an attractor. After selecting $\xi$ so as to yield an appropriate amplitude for primordial curvature perturbations, all other predictions for observable anisotropies of the cosmic microwave background radiation (CMB)---such as the spectral index, its running, and the tensor-to-scalar ratio---vary exponentially weakly with model parameters \cite{Kaiser:2013sna,Schutz:2013fua,Kaiser:2015usz,DeCross:2015uza}.

While remarkable in the context of inflationary model-building, this scenario suffers from the requirement of large values of $\xi$. The original discussion of Higgs inflation \cite{Bezrukov:2007ep} noted that the Higgs field's quartic self-coupling $\lambda$ and its nonminimal coupling to gravity $\xi$ must satisfy $\lambda / \xi^2 \sim {\cal O} (10^{-9})$ in order for the amplitude of primordial curvature perturbations to account for the temperature anisotropies in the CMB. For $\lambda \sim {\cal O} (10^{-1})$, this suggested an enormous value of the dimensionless nonminimal coupling constant, $\xi \sim {\cal O} (10^4)$. Such a large value not only raised the question of naturalness, but is also subject to many subtleties related to unitarity \cite{Burgess:2009ea,Burgess:2010zq,Barbon:2009ya,Hertzberg:2010dc,Bezrukov:2010jz,Giudice:2010ka,Lerner:2011it,Bezrukov:2012hx,Bezrukov:2013fka,Calmet:2013hia,Barbon:2015fla,Steingasser:2025txd}. 

As was quickly recognized, such unattractive values for $\xi$ can be avoided by taking into account the running of $\lambda$, whose value at high energies (relevant for early-universe inflation) is controlled by its beta function, $\beta_\lambda \equiv \partial \lambda / \partial  \ln \mu$, where $\mu$ is the renormalization group (RG) scale \cite{DeSimone:2008ei,Bezrukov:2008ej,Bezrukov:2009db,Barvinsky:2009fy,Barvinsky:2009ii,Bezrukov:2012sa,Allison:2013uaa}. For the SM Higgs field, the function $\beta_\lambda$ depends sensitively upon the couplings of the Higgs to the other SM particles, such as the Yukawa coupling for the top quark $y_{\rm t}$. Due to the relatively large value of $y_{\rm t}$, the resulting running drives $\lambda$ towards smaller values at high energies, alleviating the need for such exponentially large values of $\xi$. Upon taking into account such running of the couplings, successful inflation requires $\xi \sim {\cal O} (10^2 - 10^3)$ rather than ${\cal O} (10^4)$ \cite{Rubio:2018ogq}.

Yet incorporating the running of the SM couplings can yield its own problems for Higgs inflation. For the central values of these couplings obtained from the global averages in Ref.~\cite{ParticleDataGroup:2024cfk}, the coupling $\lambda$ turns \emph{negative} near the instability scale $\mu_I\sim 10^{11}$~GeV \cite{Steingasser:2023ugv,Detering:2024vxs}. Taking these values at face value and assuming no Beyond-the-Standard-Model (BSM) physics to alter this behavior, this running would suggest that the Higgs field's potential ``flips over'' at this scale, preventing slow-roll inflationary dynamics \cite{Rubio:2018ogq}. 

Given this behavior of $\lambda$ under RG flow, several authors introduced a variant model known as ``critical Higgs inflation'' \cite{Bezrukov:2014bra,Hamada:2014iga,Hamada:2014wna,Rubio:2018ogq}. The 95\% confidence ellipsoid of the relevant SM parameters---the top-quark mass, the Higgs mass, and the strong gauge coupling $g_s$, which affects the running of $y_{\rm t}$---contains a special point giving rise to behavior absent in the tree-level theory. Let $\mu_{\rm m}$ denote the RG scale at which all terms linear in $\ln(h/\mu)$ within the effective quartic coupling vanish, that is, where $\lambda (\mu)$ reaches its minimum. (Here $|\mathbf{H}|^2=h^2/2$ parametrizes the magnitude of the Higgs doublet.) In the vicinity of $\mu_{\rm m}$, the effective Higgs self-coupling takes the form
\begin{equation}
    \lambda_{\rm eff}(\mu_{\rm m},h)= \lambda_{\rm nl}(\mu_{\rm m}) + \lambda_{\rm ll}(\mu_{\rm m}) \ln^2 \left( \frac{h}{\mu_{\rm m}} \right) ,
    \label{lambdaeffdef}
\end{equation}
where $\lambda_{\rm nl}$ indicates the field-independent value of the quartic self-coupling at two-loop accuracy evaluated at $\mu_{\rm m}$, and $\lambda_{\rm ll}$ the coefficient of the terms quadratic in ${\rm ln} (h)$ evaluated at the same scale. 

\begin{figure*}
    \centering
    \includegraphics[width=0.8\linewidth]{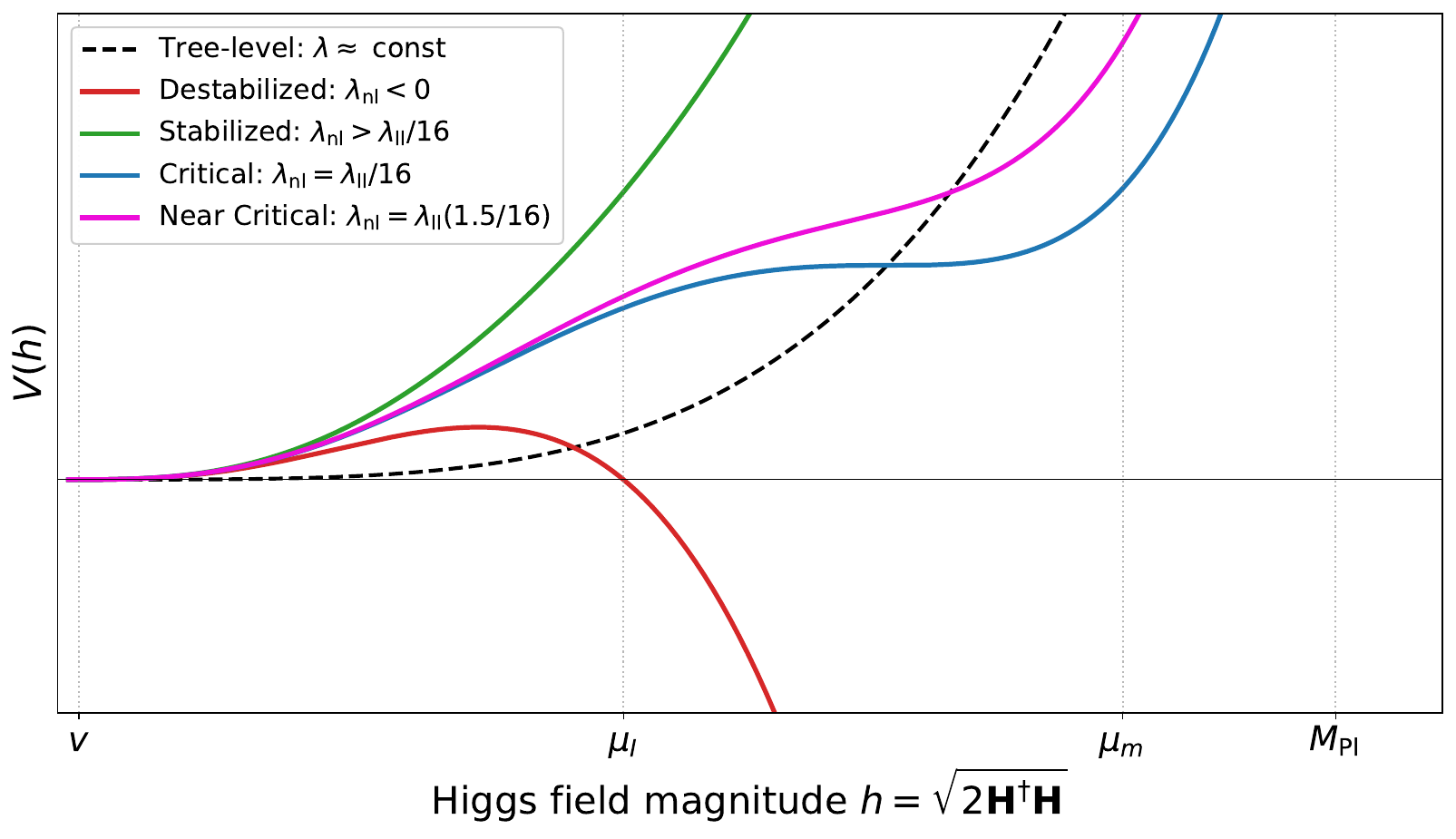}
  \caption{\justifying The effective potential for (critical) Higgs inflation in the Jordan frame for various scenarios considered here.
  \emph{Tree level} (dashed grey curve): the classical Standard Model potential $V_{\rm tree}(h)=\frac{1}{4}\lambda_{\rm SM}h^4$ is stable and monotonic. The red, green, blue, and pink curves are all variations on the two-loop-corrected Higgs potential. What distinguishes them is the behavior of the effective quartic self-coupling $\lambda_{\rm eff}$ under RG flow. \emph{Destablized} (red curve): the two-loop RG-improved potential using the central SM input values for $y_{\rm t}, g_s, m_h$, in which the running drives $\lambda_{\rm eff}(\mu)$ to negative values near the instability scale $\mu_I$. \emph{Stabilized} (green curve): the two-loop potential evaluated at SM inputs shifted (relative to those used for the red curve), within the current experimental uncertainties, chosen so that $\lambda_{\rm eff}$ remains positive throughout the relevant field range and no second minimum forms. \emph{Critical} (blue curve): A fine-tuned variant in which $m_t, m_h, g_s$ are chosen (again, within experimental uncertainties) such that the criticality condition in Eq.~(\ref{lambdanlcrit}) is satisfied. With this choice, the constant part of $\lambda_{\rm eff}$ takes its minimum value and the potential has an exact inflection point at $h = e^{-1/4}\mu_{\rm m}$. \emph{Near-critical} (pink curve): a small deformation of the critical variation with $\lambda_{\rm nl} = (\lambda_{\rm ll}/16)(1 + \delta)$, with $\delta$ typically ${\cal O} (10^{-2})$, referred to throughout this article as ``near-critical'' Higgs inflation. The near-critical scenario is our starting point for the analysis in the following sections, in which we examine whether such models can simultaneously satisfy all three of the criteria for successful Higgs inflation described in Sec.~\ref{sec:EvaluationCriteria}.}
\label{fig:Veffcritical}
\end{figure*}

At the boundary of the 95\%-confidence ellipsoid, there exists a point for which $\lambda_{\rm nl}$ and $\lambda_{\rm ll}$ satisfy the relation 
\begin{equation}
	 \lambda_{\rm nl} = \lambda_{\rm nl,crit} \equiv  \lambda_{\rm ll}/16,
     \label{lambdanlcrit}
\end{equation}
for which the effective quartic coupling remains positive and the RG-improved effective potential for the Higgs field $h$ develops an inflection point. See Fig.~\ref{fig:Veffcritical}.

The plateau in the effective potential featuring such an inflection point has long been known to be capable of giving rise to successful slow-roll inflation, with a value of the nonminimal coupling $\xi \sim {\cal O} (30)$ \cite{Bezrukov:2014bra}. Moreover, such a model could account for the measured CMB anisotropies while also producing peaks in the power spectrum of primordial curvature perturbations at much shorter length scales, which could produce detectable gravitational waves and possibly primordial black holes \cite{Ezquiaga:2017fvi,Bezrukov:2017dyv}. 

Our goal in this work is to evaluate critical Higgs inflation in the light of three distinct criteria: (1) consistency with the running of SM couplings from the electroweak (EW) scale, at which they are well measured, to the energy scales relevant for inflation; (2) predictions for cosmological observables, such as CMB anisotropies;  and (3) self-consistent description of the inflationary dynamics at high energies while \emph{also} reproducing the familiar Higgs-sector phenomenology at the EW scale.

As we demonstrate here, there do not exist \emph{any} regions of parameter space within which all three criteria can be satisfied simultaneously. Models of the very early universe that are governed by an action like that for critical Higgs inflation can readily provide excellent agreement with a host of cosmological observables, but only at the cost of self-consistency when treated as effective descriptions of EW-scale physics: the theory as defined near the EW scale violates perturbative unitarity at large momenta, necessitating a UV completion to restore unitarity above some energy $E_{\rm max}$. As we find here, across the relevant regions of parameter space, $E_{\rm max}$ is small enough that it should have a significant effect on the inflationary dynamics. While it \textit{would} be possible that this effect becomes subdominant in the inflationary regime, any such statement would represent an additional assumption about the details of the UV completion.

The remainder of this article is structured as follows. In Sec.~\ref{sec:EvaluationCriteria}, we briefly describe the three criteria we use to evaluate critical Higgs inflation. Sec.~\ref{sec:NearCriticalHiggs} introduces the effective action for critical Higgs inflation, including formulations in both metric gravity and Palatini (or metric-affine) gravity. In Sec.~\ref{sec:HiggsEWscale} we use the most recent compilations of EW-scale measurements to determine parameters within the SM Higgs sector as well as their running to higher energies under RG flow, relevant for early-universe inflation. We then discuss the realization of slow-roll inflation in this model in Sec.~\ref{sec:Inflation} and briefly review the CMB observables that the model should reproduce. In Sec.~\ref{sec:Unitarity} we discuss implications of perturbative unitarity for self-consistent regions of parameter space, within which predictions from the model should remain insensitive to higher-order BSM contributions to the effective action. 

In the light of these three distinct criteria, we present numerical results in Sec.~\ref{sec:NumericalResults} for the behavior of model predictions from critical Higgs inflation, in both metric and Palatini formulations, across allowable regions of parameter space. Brief concluding remarks follow in Sec.~\ref{sec:Discussion}. Several appendices include additional details regarding the criticality condition, RG flow of SM parameters, implications of conformal transformations for both metric and Palatini formulations, and details of our numerical implementation.

\section{Three Criteria}
\label{sec:EvaluationCriteria}

We evaluate critical Higgs inflation according to three distinct criteria, which we briefly describe in this section.

\subsection{Criterion 1: Consistency with EW-Scale Measurements}
\label{subsec:Criterion1}

The first criterion concerns model ingredients: is the inflaton identified with the same scalar fields responsible for EW symmetry-breaking in the SM? This criterion concerns both the \emph{functional form} of the effective action and the \emph{allowable regions of parameter space}. For functional form, we require that the interaction potential only include Higgs-sector features (including loop corrections) that appear within the SM. Identifying the inflaton with the SM Higgs field, in turn, significantly constrains the relevant parameter space, since SM couplings have been measured precisely in collider experiments near the EW scale, and the structure of the SM determines the $\beta$-functions under which such couplings vary under RG flow.

Insisting that inflation be driven by the SM Higgs sector requires the model to reproduce the relevant EW-scale measurements. Whereas it is well-understood that the effect of any nonminimal coupling of the Higgs field to gravity is negligible in this regime, the same is not true for the parameters $\lambda_{\rm nl}$, $\lambda_{\rm ll}$, and $\mu_{\rm m}$ defined in Eq.~(\ref{lambdaeffdef}), which can be obtained directly from the values of the SM couplings at the EW scale. Thus, a straightforward procedure to obtain the region of parameter space consistent with EW-scale measurements is to identify the corresponding parameter regime at the EW scale and then map this regime to high energies by integrating the coupled $\beta$-functions for each SM parameter. 

This approach might, however, be spoiled by the strong sensitivity of the RG running in the vicinity of a critical point, which would allow small deviations to lead to a qualitative change in the shape of the potential \cite{Steingasser:2023ugv}. Indeed, we find that the parameter $\lambda_{\rm nl}$ is particularly sensitive to small changes in the RG running near the critical point identified in Eq.~(\ref{lambdanlcrit}), which could, for instance, arise from some novel BSM physics. As described in more detail in Sec.~\ref{sec:HiggsEWscale}, we therefore treat $\lambda_{\rm nl}$ as an essentially free parameter while calculating SM-compliant regions for $\lambda_{\rm ll}$ and $\mu_{\rm m}$. In particular, we parameterize $\lambda_{\rm nl}=(1+\delta)\lambda_{\rm nl, crit}$, in terms of some new free parameter $\delta$, with $\vert \delta \vert \ll 1$. This amounts to a \textit{extension} of the near-critical region of parameter space permitted by the pure SM.

\subsection{Criterion 2: Compliance with CMB Observables}
\label{subsec:Criterion2}

The second criterion concerns predictions for cosmological observables: does critical Higgs inflation predict features of the CMB in close agreement with precision measurements?

Unlike EW-scale measurements, model predictions for cosmological observables depend on the nonminimal coupling $\xi$ as well as on $\lambda_{\rm ll},\ \mu_{\rm m}$, and $\delta$. In particular, the amplitude of primordial curvature perturbations---which is directly related to the amplitude of temperature anisotropies measured in the CMB---is strongly sensitive to $\xi$. We find that for a given set of values $\{ \lambda_{\rm ll}, \mu_{\rm m},\delta \}$, one may always find a value of $\xi$ such that predictions for all CMB observables match the latest measurements. Such a possibility has long been recognized when considering the original, tree-level model of Higgs inflation \cite{Bezrukov:2007ep,Rubio:2018ogq,Kaiser:2013sna,Schutz:2013fua,Kaiser:2015usz}. We demonstrate here that this feature persists for the loop-corrected effective action near the critical point, in both the metric-gravity and Palatini formulations.

Symbolically, the different constraints therefore enter as
\begin{align}
     \{\underbrace{\lambda_{\rm ll},\mu_{\rm m}}_{\rm SM},\underbrace{\delta}_{\rm near-criticality},\underbrace{\xi}_{\rm CMB}\}.
\end{align}
In this way, combining precision EW-scale measurements, imposing near-criticality at inflationary energy-scales, and requiring compliance with CMB observables amounts to fully determining the model parameters.

\subsection{Criterion 3: Self-Consistency in both EW-Scale and Inflationary Regimes}
\label{subsec:Criterion3}

The third criterion concerns self-consistency: does the \emph{same} effective action provide a close fit to cosmological observables while \emph{also} reproducing the well-studied EW-scale phenomenology, without requiring the addition of new, higher-order BSM terms to which model predictions might become sensitive? 

This criterion arises from considering the behavior of scattering amplitudes at various energy scales. Previous works have established that upon including the nonminimal coupling $\xi {\bf H}^\dagger {\bf H} R$ in the effective action, the most stringent constraints on perturbative unitarity arise from $2 \rightarrow N$ scattering of Higgs particles. (See Ref.~\cite{Steingasser:2025txd} and references therein.) 

If one considers such scattering amplitudes \emph{during inflation}, by expanding the effective action in perturbations around the nontrivial Higgs condensate, then one finds a breakdown of perturbative unitarity at center-of-mass energy \cite{Steingasser:2025txd}
\begin{equation}
E \gtrsim E_{\rm max}^{\rm infl} \simeq \frac{ 4 \sqrt{\pi} \, M_{\rm Pl} } {\sqrt{\xi}} 
\label{Emaxinfl}
\end{equation}
in the case of metric gravity, where $M_{\rm Pl} \equiv 1 / \sqrt{ 8 \pi G} = 2.43 \times 10^{18} \, {\rm GeV}$ is the reduced Planck mass. For $\xi \lesssim {\cal O} (50)$---as is typical for models like critical Higgs inflation in the metric-gravity formulation---this implies that an effective action that only includes the SM Higgs sector provides a self-consistent description of inflationary dynamics.

However, models of Higgs inflation (including critical Higgs inflation) are intended to \emph{also} provide a self-consistent description of EW-scale phenomena. This implies that \emph{the same} effective action must remain well-behaved even when treating phenomena well below the inflationary energy scale. When calculating $2 \rightarrow N$ scattering amplitudes among Higgs particles around the low-energy EW vacuum, perturbative unitarity breaks down at center-of-mass energies \cite{Steingasser:2025txd}
\begin{align}
    E\gtrsim E_{\rm max}^{\rm EW}\simeq \Bigg\{ \begin{array}{ll}
       2.05 \cdot M_{\rm Pl} / \xi  & \quad{\rm metric\ gravity} \\ 
        4.94 \cdot M_{\rm Pl} /\sqrt{\xi}   & \quad{\rm Palatini\ gravity} \end{array}
    \, .
\label{EmaxEW}
\end{align}
Preserving unitarity in such scattering processes would require some new physics to modify the effective action above some scale $\Lambda \lesssim E_{\rm max}^{\rm EW}$. 

Below this scale, the impact of such (as-yet unknown) new physics could be described via a tower of higher-dimensional BSM operators of the form
\begin{align}
    \frac{\Delta \mathcal{L} }{\sqrt{-g}}\supset \frac{C_H}{\Lambda^2} |\mathbf{H}|^6 + \frac{C_{H \Box}}{\Lambda^2}|\mathbf{H}|^2 |\Box \mathbf{H}|^2+... \, ,
    \label{BSMterms}
\end{align}
where $C_H$ and $C_{H\Box}$ are Wilson coefficients. Thus, for generic Wilson coefficients, these additional terms would have a nonnegligible impact on the inflationary dynamics \emph{if} $\vert {\bf H} \vert^2 \sim \Lambda^2$. In such cases, model predictions for CMB observables would become sensitive to the (unknown) higher-order BSM terms, parameterized by the Wilson coefficients $C_i$. In other words, model predictions for cosmological observables would become sensitive to unknown UV physics.

One could try to avoid the need to incorporate higher-order BSM terms such as those in Eq.~(\ref{BSMterms}) by raising the energy scale $E_{\rm max}^{\rm EW}$ in Eq.~(\ref{EmaxEW}). Such a shift could be accomplished by changing the running of the SM couplings under RG flow---but such a shift would require some \emph{other} BSM physics, in order to alter the SM $\beta$-functions. Alternatively, one might insist that the properties of the BSM physics are such that it decouples from the SM Higgs during inflation, thus leaving its inflationary dynamics unaltered. In either scenario, predictions for cosmological observables from models like (critical) Higgs inflation would become sensitive to unknown BSM physics.

As discussed in further detail in Sec.~\ref{sec:Unitarity}, our third criterion therefore concerns a comparison of some appropriate measure of the inflaton field during inflation to the scale $\Lambda \lesssim E_{\rm max}^{\rm EW}$, for both metric and Palatini formulations of critical Higgs inflation. 

\section{Effective action for critical Higgs inflation}
\label{sec:NearCriticalHiggs}

In this section we introduce the effective potential for the SM Higgs doublet near its critical point at high energies relevant for early-universe inflation. The scalar degrees of freedom of the SM EW sector consist of a complex doublet,
\begin{equation}
    {\bf H} = \left( \begin{array}{c} H^+ \\ H^0 \end{array} \right) .
\label{Hdoublet}
\end{equation}
The complex fields $H^+$ and $H^0$ may be further decomposed into (real) scalar degrees of freedom,
\begin{equation}
\begin{split}
H^+ &= \frac{1}{ \sqrt{2}} \left( \phi^1 + i \phi^2 \right) , \\
H^0 &= \frac{1}{ \sqrt{2}} \left( h + i \phi^3 \right) ,
\end{split}
\label{Hchi}
\end{equation}
where $h\equiv \phi^0$ is the Higgs scalar and the $\phi^a$ ($a = 1, 2, 3$) are the Goldstone modes. (It is convenient to group these as $\phi^I = (h, \phi^a)$, where the index $I = 0, 1,2 , 3$.) Within renormalizable $R_\xi$ gauges, all four scalar degrees of freedom remain in the spectrum at high energies \cite{Weinberg:1996kr}. 

Above the symmetry-breaking scale, the effective potential depends only on the combination ${\bf H}^\dagger {\bf H}$. For any model with ${\cal N}$ real-valued scalar fields that respects an SO(${\cal N}$) symmetry, the background dynamics depend only on three initial conditions: the initial magnitude and initial velocity along the radial direction in field space, and the initial velocity perpendicular to the radial direction. Without loss of generality, we may therefore exploit the gauge symmetry of the EW sector to consider the dynamics of just two scalar degrees of freedom, $h(t)$ and $\phi(t)$, with $\phi (0) = 0$. The remaining dependence of the dynamics on $\dot{\phi} (0)$ distinguishes the behavior of the model from a genuinely single-field model, in which one neglects (or gauges away) the Goldstone modes altogether. Finally, in a cosmological context, any initial kinetic energy in the Goldstone modes redshifts away rapidly, $\dot{\phi}^2 \sim a^{-6} (t)$; generically, the system settles into a ``single-field attractor'' within ${\cal O}(1)$ e-folds after the onset of inflation \cite{Greenwood:2012aj} (see also Refs.~\cite{Kaiser:2012ak,Kaiser:2013sna,Schutz:2013fua,Kaiser:2015usz,DeCross:2015uza}). Following this initial transient, we may use the gauge symmetry to perform a rotation in field space, such that 
\begin{equation}
{\bf H}^\dagger {\bf H} \rightarrow \frac{1}{2} h^2,
\label{Hmagnitudeh}
\end{equation}
or, effectively, $\vert {\bf H} \vert \rightarrow h / \sqrt{2} $. The resulting dynamics thus mimic those of the unitary gauge. Given the analysis in Ref.~\cite{Irges:2017ztc}, these transformations are consistent with accurate evaluation of loop corrections as well. 

Well above the EW symmetry-breaking scale $v = 246$ GeV, the effective potential for the Higgs sector then may be written as
\begin{equation}
    V_{\rm eff} (h, \mu)= \frac{1}{4} \lambda_{\rm eff}(\mu, h)\,  h^4 ,
    \label{Veff1}
\end{equation}
where $h$ contains a wave function normalization factor, $h \to e^{\Gamma (\mu)} h$. The effective quartic self-coupling, incorporating corrections up to two-loop accuracy, takes the form
\begin{equation}
    \lambda_{\rm eff}(\mu, h) \simeq \lambda_{\rm nl} (\mu) + \lambda_{\rm l} (\mu) \,\ln \left( \frac{h}{\mu} \right)  + \lambda_{\rm ll} (\mu) \,\ln^2 \left( \frac{h}{\mu} \right) . \label{eq:lambdaexp}
\end{equation}
For generic parameters, the coefficient of each power of $\ln(h/\mu)$ is suppressed by the corresponding order in the loop expansion.\footnote{We discuss how these parameters can be expressed in terms of the corresponding $\beta$ functions in Appendix~\ref{appendixBetaFunctions}.} For such generic parameters, this usually implies that the dynamics of the Higgs field is to leading order controlled by the tree-level contribution $\lambda_{\rm eff}\simeq \lambda_{\rm nl}$. This corresponds to the original Higgs inflation scenario \cite{Bezrukov:2007ep}, which has been explored extensively in the literature~\cite{Rubio:2018ogq}. 

Whereas models of Higgs inflation have typically been studied in the limit $\lambda_{\rm eff} \simeq \lambda_{\rm nl}$, the effective potential of Eq.~(\ref{Veff1}) displays interesting behavior near its {\it critical point}, which marks a transition between two quantum phase transitions. This behavior can be understood through the RG running of $\lambda$: due to a large negative contribution of the top quark to $\beta_\lambda$, $\lambda$ runs toward smaller values at higher energies. Depending on the values of the SM couplings near the EW scale, current measurements and their uncertainties allow for two possibilities: for a vast majority of parameter values allowed at 95\% likelihood, $\lambda$ turns negative at the \textit{instability scale} $\mu_I$, causing the potential to ``flip over'' at $h\sim \mu_I$. On the other hand, for a small region within the same 95\% likelihood contours for the SM parameters, the coupling $\lambda$ remains positive, with the potential growing indefinitely at larger field values. The transition between these two phases is marked by a \textit{critical potential} containing an inflection point \cite{Steingasser:2023ugv}.

It is easy to see that such critical behavior requires $\lambda$ to decline to some small but positive minimum value at scale $\mu_{\rm m}$, above which $\lambda$ grows to larger values. This immediately suggests that one consider the effective coupling $\lambda_{\rm eff}$ up to second order in the logarithmic expansion of Eq.~\eqref{eq:lambdaexp}. In order for the term quadratic in logarithmic corrections to have a qualitative impact on the behavior of the potential, one requires $\lambda_{\rm nl}\sim \lambda_{\rm ll}\sim\mathcal{O}({\rm two-loop})$. Moreover, expanding around the minimum suggests that one consider the potential at the fixed RG scale $\mu_{\rm m}$, such that the relevant potential becomes
\begin{equation}
    V(h, \mu_{\rm m}) \equiv \frac{1}{4} \left( \lambda_{\rm nl} (\mu_{\rm m})   + \lambda_{\rm ll} (\mu_{\rm m}) \,\ln^2 \left( \frac{h}{\mu_{\rm m}} \right) \right) h^4.
    \label{Vcritical1}
\end{equation}
Moving forward, we will drop the explicit dependence on $\mu_{\rm m}$ and assume that all scale-dependent quantities are evaluated at this scale.

In addition to the effective potential for the Higgs field, models of Higgs inflation also incorporate a nonminimal coupling of the Higgs doublet to gravity, typically of the form
\begin{align}
    \mathcal{L} \supset \sqrt{-\tilde{g}} f(h) \tilde{R} ,
\end{align}
where $\tilde{g}$ and $\tilde{R}$ denote the spacetime metric and its Ricci scalar. The function $f$ is a scalar-valued function of the Higgs doublet ${\bf H}$, whose gauge invariance implies that the nonminimal coupling depends only on the magnitude $h$. One of the most common choices for this function is 
\begin{align}
    f(h)=\frac{1}{2}\left(M_{\rm Pl}^2+ \xi h^2 \right).
    \label{eq:NMcoup}
\end{align}
The first term, involving the reduced Planck mass $M_{\rm Pl}$, multiplies the usual kinetic term for the metric $\tilde{g}$. The second term represents the nonminimal coupling between the Higgs field and gravity. This specific form of the coupling is of particular interest, since it is generated via loop corrections even if one were to set $\xi=0$ at tree-level \cite{Birrell:1982ix,Buchbinder:1992rb,Parker:2009uva}.

The effect of the nonminimal coupling on the potential can be conveniently understood by performing a conformal transformation to the ``Einstein frame,'' in which
\begin{align}\label{eq:conftransf}
    g_{\mu \nu}(x)= \Omega^2(x) \, \tilde{g}_{\mu \nu}(x), \>\, {\rm where} \>\, \Omega^2(x)=1+ \xi \frac{h^2}{M_{\rm Pl}^2}.
\end{align}
In this new frame, the potential takes the form (see, e.g., Ref.~\cite{Kaiser:2015usz})
\begin{align}\label{eq:Vrescaled}
    V_E(h, \mu)= \frac{V(h, \mu)}{\left(1+ \xi h^2/M_{\rm Pl}^2 \right)^2}.
\end{align}
While this transformation yields the familiar Einstein-Hilbert action for the metric tensor, it also induces a nontrivial field-space metric for the components of the Higgs doublet \cite{Kaiser:2010ps}
\begin{equation}
\begin{split}
    \tilde{g}^{\mu \nu}\partial_\mu \mathbf{H}^\dagger \partial_\nu \mathbf{H} &= \frac{1}{2} \tilde{g}^{\mu \nu} \delta_{IJ} \partial_\mu \phi^I \partial_\nu \phi^J \\
    &\longrightarrow \frac{1}{2}  g^{\mu \nu} {\cal G}_{IJ} (\phi^K) \, \partial_\mu \phi^{I} \partial_\nu \phi^J,
    \label{kineticsector}
\end{split}
\end{equation}
where
\begin{equation}
     {\cal G}_{IJ} (\phi^K) = \frac{M_{\rm Pl}^2}{M_{\rm Pl}^2+ \xi h^2}\left[ \delta_{IJ} + 6 \sigma \xi^2 \frac{\delta_{IK} \delta_{JL} \phi^K \phi^L}{M_{\rm Pl}^2+ \xi h^2} \right]. \label{eq:GIJ}
\end{equation}
The field-space indices $I, J = 0, 1, 2, 3$, where, as noted below Eq.~(\ref{Hchi}), $\phi^0 = h$ and the remaining components represent the real-valued Goldstone modes.

In Eq.~\eqref{eq:GIJ}, the choice $\sigma=1$ corresponds to metric gravity, whereas $\sigma=0$ represents Palatini gravity.\footnote{As discussed in Appendix~\ref{App:ConformalTransformation}, the term in Eq.~(\ref{eq:GIJ}) that is proportional to $\sigma$ arises from the transformation of the connection $\Gamma^\mu_{\nu\lambda}$ under the conformal transformation of Eq.~(\ref{eq:conftransf}), and therefore  the presence or absence of this term depends upon the relationship between the metric $g_{\mu\nu}$ and the connection $\Gamma^\mu_{\nu\lambda}$.} Finally, the conformal transformation of Eq.~\eqref{eq:conftransf} also affects the masses of the remaining SM particles, and through them the loop corrections to the potential. As pointed out in Ref.~\cite{Bezrukov:2009db}, this amounts to the replacement
\begin{align}
    \ln \left(\frac{h}{\mu_{\rm m}} \right)\to \ln \left(\frac{h}{\sqrt{M_{\rm Pl}^2+\xi h^2}}\cdot \frac{M_{\rm Pl}}{\mu_{\rm m}} \right).
\end{align}
Together with the overall rescaling factor of the potential in Eq.~\eqref{eq:Vrescaled}, this suggests introducing an auxiliary variable $z$, defined as
\begin{align}\label{eq:zdef}
    z (h) \equiv  \frac{h}{\sqrt{M_{\rm Pl}^2 + \xi h^2 }} \cdot \frac{M_{\rm Pl}}{\mu_{\rm m}} .
\end{align}
In terms of this new variable, the Einstein-frame potential near its critical point can be brought to the simple form
\begin{align}\label{eq:VE}
    V_{E}(h, \mu_{\rm m})=\frac{1}{4} &\left[ \lambda_{\rm nl}+ \lambda_{\rm ll} \cdot  \ln^2 \left(z(h)\right)  \right] \cdot  z^4(h) \cdot \mu_{\rm m}^4,
\end{align}
which resembles the structure of the potential in Eq.~(\ref{Vcritical1}).

For a potential of the form in Eq.~\eqref{eq:VE}, it is straightforward to demonstrate that the critical point occurs for $\lambda_{\rm nl, crit} = \lambda_{\rm ll} / 16$, as in Eq.~(\ref{lambdanlcrit}), as we briefly review in Appendix~\ref{App:Criticality}. At the exact critical point corresponding to the condition in Eq.~(\ref{lambdanlcrit}), the potential in Eq.~(\ref{eq:VE}) has an inflection point at $z = e^{-1/4}$. We note, however, that Eq.~(\ref{eq:zdef}) implies an upper bound on $z$:
\begin{align}
    z (h) 
    <\frac{M_{\rm Pl}}{\sqrt{\xi}\, \mu_{\rm m} } ,
\end{align}
which in turn implies a necessary condition for an inflection point to form for any finite value of the Higgs field,
\begin{gather}
    \mu_{\rm m} < \frac{e^{1/4} \, M_{\rm Pl}}{\sqrt{\xi}} .
\label{inflationpointbound}
\end{gather}
As we will see in Sec.~\ref{sec:Inflation}, scenarios of critical Higgs inflation exist for which $\mu_{\rm m}$ satisfies the bound in Eq.~(\ref{inflationpointbound}).

\section{Implications of EW-scale measurements}
\label{sec:HiggsEWscale}

One of the strongest motivations behind Higgs inflation is the possibility to relate the inflationary dynamics to measurements at the EW scale. In the special case of near-critical Higgs inflation, the parameters $\mu_{\rm m}$ and $\lambda_{\rm ll}$ in the potential can be related to the RG trajectory of $\lambda$ through
\begin{align}
    \beta_\lambda (\mu_{\rm m})=0, \quad \lambda_{\rm ll} (\mu_{\rm m}) = \frac{1}{2} \left( \mu \frac{ \partial \beta_\lambda}{\partial \mu} \right) \Bigg\vert_{\mu = \mu_{\rm m}} .
\end{align}
For completeness, we provide a derivation of these relations in Appendix~\ref{appendixBetaFunctions}, together with the relevant $\beta$-functions as listed, e.g., in Refs.~\cite{Buttazzo:2013uya,Khoury:2021zao}.

\begin{figure*}[t]
    \centering                                
    \includegraphics[width=.66\textwidth]{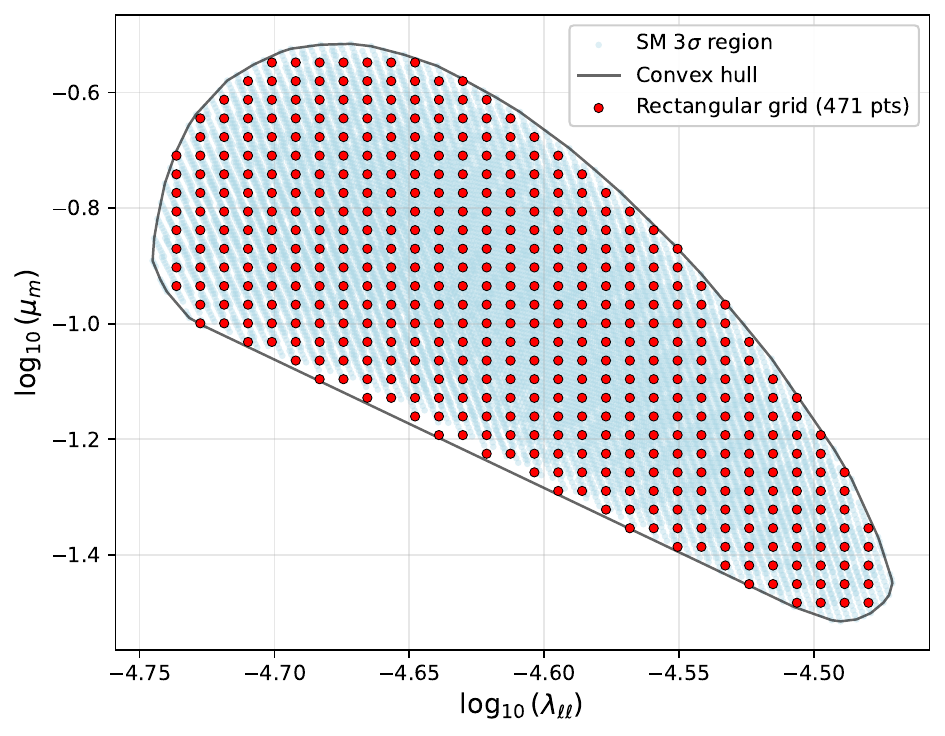}
    \caption{\justifying The region of the $(\log_{10}\lambda_{\rm ll},\log_{10}\mu_{\rm m})$ plane (light blue) that is consistent within $3 \sigma$ with precision EW-scale measurements, enclosed by its convex hull (black). This region is the image in the log-space $(\lambda_{\rm ll},\mu_{\rm m})$ plane of the 3$\sigma$ ellipsoid on the SM inputs $(m_t, m_h, g_s)$ derived from the EW precision data propagated through the SM RG running to the inflationary scale, as described in Sec.~\ref{sec:HiggsEWscale}. Superposed upon this region is a grid of 471 points (red dots) obtained from a uniform $32 \times 32$ rectangular grid spanning the hull's axis-aligned bounding box, masked to the hull interior.
    }
    \label{fig:chickenpoxplot}              
\end{figure*}

As input at the EW scale, we take the values of the relevant couplings at the top-quark mass scale $\mu = M_t=173.1$, as calculated recently in Ref.~\cite{Detering:2024vxs} based on Ref.~\cite{ParticleDataGroup:2024cfk}. Including the relevant error bars,\footnote{We are grateful to the authors of Ref.~\cite{Detering:2024vxs} for providing these numerical values.} 
these are given by
\begin{gather}
    g_1= 0.349946,\ g_2 = 0.652825,\\ 
    g_3 = 1.1650 \pm 0.0043, \
\lambda = 0.12987 \pm 0.00024,  \nonumber\\ 
y_{\rm t} = 0.9373 \pm 0.0048, \ y_b = 0.02,\ y_\tau = 0.01.  \nonumber
\end{gather}
Using this input, we construct the $68.3\%$ ($1  \sigma$), $95.4\%$ ($2  \sigma$), and $99.7\%$ ($3 \sigma$) confidence ellipsoids in $\lambda$-$g_3$-$y_{\rm t}$-space, keeping fixed the other couplings. To infer the corresponding ranges in $\mu_{\rm m}$-$\lambda_{\rm nl}$-$\lambda_{\rm ll}$-space, we cover our input error ellipsoid with a homogeneous grid of $\sim 10^4$ points and integrate the $\beta$-functions for each parameter. We present our exact results in Table~\ref{table:confidence}. Due to the nonlinear nature of the RG running close to the critical point, the original error ellipsoid is distorted while being mapped onto $\mu_{\rm m}$-$\lambda_{\rm nl}$-$\lambda_{\rm ll}$-space. In Fig.~\ref{fig:chickenpoxplot}, we provide the projection of this space onto the $\mu_{\rm m}$-$\lambda_{\rm ll}$-plane for the $3\sigma$ confidence interval.

\begin{table}
\centering
\begin{ruledtabular}
\begin{tabular}{ c c c c c } 
 & Central & $68.3\%$  & $95.4\%$ & $99.7\%$ \\
\hline
$\mu_{\rm m}/M_{\rm Pl}$ & $0.12  $ & $[0.06,0.2] $&  $[0.04,0.2]$ & $[0.03,0.3]$ \\
\hline
$\lambda_{\rm nl}$ & $-0.01$  & $[-0.02,0.00043 ]$ & $[-0.03,0.006]$ & $[-0.03,0.01]$ \\
\hline
$\lambda_{\rm ll} \cdot 10^5$ & $2.4$ & $[2.1,2.9]$ & $[1.9,3.1]$ & $[1.8,3.4]$ \\
\end{tabular}
\end{ruledtabular}
\caption{\justifying The range of values of the potential parameters $\mu_{\rm m}$, $\lambda_{\rm nl} (\mu_{\rm m})$, and  $\lambda_{\rm ll} (\mu_{\rm m})$ at several confidence levels in the EW-scale input data. Across a majority of parameter space, we find $\lambda_{\rm nl}<0$, corresponding to the Higgs potential flipping over. This behavior, however, is subject to theoretical subtleties related to the vicinity of $\mu_{\rm m}$ to a quantum critical point. First, this manifests through the strong sensitivity of $\lambda_{\rm nl}$ to the EW input, which would also extend to possible BSM effects on the running. Second, it could be understood as hinting at the presence of some dynamical mechanism responsible for setting $\lambda$ in the UV, which would favor values close to the critical value $\lambda_{\rm nl}=\lambda_{\rm ll}/16$~\cite{Steingasser:2023ugv}. Both of these effects are absent in this table, but are crucial for our analysis in Sec.~\ref{sec:NumericalResults}.}
\label{table:confidence}
\end{table}

Whether or not inflation is possible in the first place hinges on the sign of $\lambda_{\rm nl} (\mu_{\rm m})$, since $\lambda_{\rm nl}<0$ would correspond to the potential flipping over, causing the scalar field to roll towards larger field values and ultimately the Universe to crunch. In agreement with earlier investigations~\cite{Buttazzo:2013uya,Steingasser:2023ugv,Detering:2024vxs,Hiller:2024zjp}, our analysis finds this behavior in a majority of parameter space: for the $1\sigma$-confidence ellipsoid, we find that $99.6\%$ of the parameter-space points covered in our scan yield $\lambda_{\rm nl} (\mu_{\rm m})<0$, including the central value. Within the $2\sigma$- and $3\sigma$-confidence ellipsoids, we find this behavior for $90.7\%$ and $82.5\%$ of the parameter-space points, respectively. 

Earlier works~\cite{Buttazzo:2013uya,Gonderinger:2009jp,Elias-Miro:2012eoi,Chen:2012faa,Rodejohann:2012px} have pointed out that this conclusion could be dramatically altered through the effect of BSM physics capable of restoring vacuum stability. These authors have suggested to quantify such effects through the introduction of a threshold correction to the quartic coupling,
\begin{align}
    \lambda_{\rm nl,SM} \to  \lambda_{\rm nl,SM} + \Delta \lambda_{\rm nl} \equiv  \lambda_{\rm nl}.
\end{align}
For the purpose of our analysis, we combine the possibility of such an effect with the SM error bars by treating the overall $\lambda_{\rm nl}$ effectively as a free parameter.\footnote{Whereas in principle such BSM effects could also influence the parameters $\lambda_{\rm ll} (\mu_{\rm m})$ and $\mu_{\rm m}$, we expect that the impact on these parameters would be less significant than on $\lambda_{\rm nl}$, and likely comparable to the measurement uncertainties at the EW scale that we have already incorporated within our analysis. Therefore we only consider possible effects on $\lambda_{\rm nl} (\mu_{\rm m})$ in our analysis.}
In particular, near the critical point, we parameterize
\begin{align}
    \lambda_{\rm nl}= \lambda_{\rm nl,crit}(1+\delta), \>\, {\rm where} \>\, \lambda_{\rm nl,crit}=\lambda_{\rm ll}/16.
    \label{deltaparameter}
\end{align}
We focus on the case $\delta >0$, corresponding to a monotonically growing potential.

In our main analysis, EW-scale measurements will manifest through the parameters $\{\mu_{\rm m},\lambda_{\rm ll},\delta\}$. In order to focus our scan on near-critical behavior, we will choose $0 < \delta \ll 1$, ignoring any correlation with $\mu_{\rm m}$ and $\lambda_{\rm ll}$. For the latter, we consider a grid of 471 points obtained from a uniform $32 \times 32$ rectangular grid of values overlaid on the SM $3\sigma$-confidence region, projected into the $(\mu_{\rm m}, \lambda_{\rm ll})$ plane. See Fig.~\ref{fig:chickenpoxplot}. Throughout the remainder of this article, we will refer to this region as the SM $3 \sigma$ region. Neglecting the correlation between the freely chosen perturbation away from exact criticality, $\delta$, and the other parameters amounts to a strict \textit{extension} of the relevant parameter space to include BSM scenarios or account for systematic errors, as discussed in the previous paragraph. We will show that even across this extended parameter space, models of critical Higgs inflation cannot simultaneously satisfy the three criteria enumerated in Sec.~\ref{sec:EvaluationCriteria}. A fortiori, neither can the baseline model with purely SM EW input parameters. 

Eq.~\eqref{deltaparameter} implies that near-critical behavior requires a deviation from the usual power counting of the loop expansion, in which $\lambda_{\rm nl}$ would be expected to be dominated by a tree-level part, while $\lambda_{\rm ll}$ only emerges at the two-loop level. Both in the pure SM as well as in our extension, which allows for threshold effects ($\lambda_{\rm nl}\to \lambda_{\rm nl}+\Delta \lambda$), the terms $\lambda_{\rm nl}$ and $\lambda_{\rm ll}$ become comparable due to $\lambda_{\rm l} (\mu_{\rm m}) \simeq 0$, leaving $\lambda_{\rm nl} \sim {\cal O} (\text{two-loop})$. It is then easy to see that these specific conditions do \textit{not} spoil the loop expansion more generally, given that the logarithms remain sufficiently small, which is assured as long as $z=\mathcal{O}(0.1-10)$, where $z (h)$ is defined in Eq.~(\ref{eq:zdef}).

\section{Inflationary Dynamics and Cosmological Observables}
\label{sec:Inflation}

In this section we consider inflationary dynamics for models of Higgs inflation near the critical point, for both the metric and Palatini formulations of gravity. 

As noted above Eq.~(\ref{Hmagnitudeh}), at energy scales far above the EW symmetry-breaking scale, and following a possible initial transient, the background dynamics of the multifield system evolve along an effectively single-field trajectory within field space, which we may select to be the $\phi^0 = h$ direction. The effectively single-field dynamics persist in these models because neither $V_E (h)$ nor ${\cal G}_{IJ} (\phi^I)$ include sufficiently large gradients to push the system away from evolving along the radial direction in field space. In the absence of any such turning among the background fields---at least during the final 60 e-folds of inflation---the system evolves with $\phi^I \sim \dot{\phi}^I \sim 0$ for $I \geq 1$ \cite{Greenwood:2012aj,Kaiser:2012ak,Kaiser:2013sna,Schutz:2013fua,Kaiser:2015usz,DeCross:2015uza}.

\begin{figure*}[t]
    \centering
    \includegraphics[width=\linewidth]{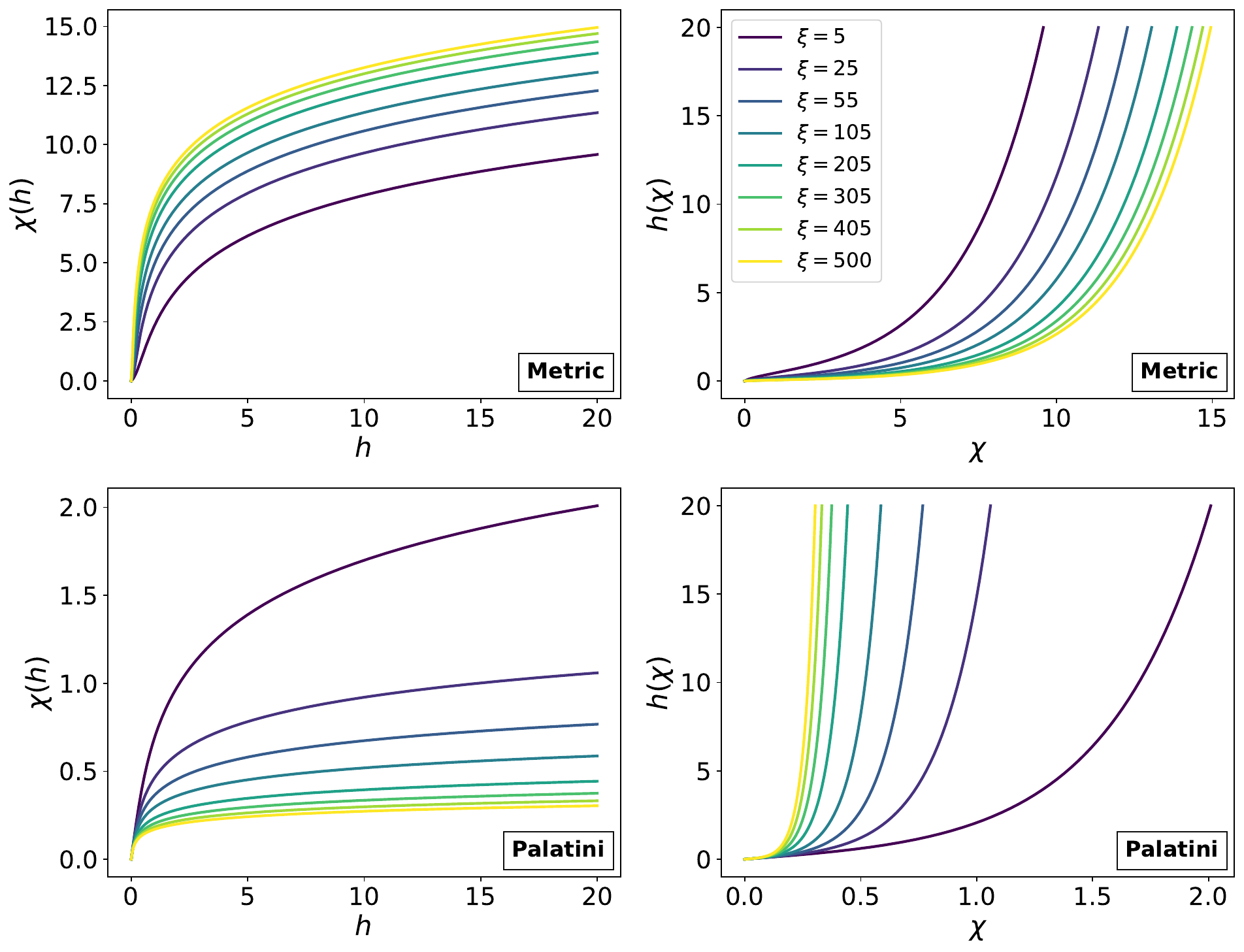}
    \caption{\justifying Transformations between the Higgs field $h$ and the canonically normalized field $\chi (h)$ in both the metric ({\it top}) and Palatini ({\it bottom}) formulations, based on Eqs.~(\ref{chimetric})--(\ref{chiPalatini}). All fields are shown in units of $M_{\rm Pl}$.
    }
    \label{fig:HiggsChiMaps}
\end{figure*}

Within the single-field attractor, the relevant component of the field-space metric in Eq.~(\ref{eq:GIJ}) is simply ${\cal G}_{hh}$, and we may introduce a single real-valued, canonically normalized scalar field $\chi$ in the Einstein frame via the rescaling \cite{Kaiser:2010ps}
\begin{equation}
    \frac{ d\chi}{dh} = \frac{1}{\sqrt{ {\cal G}_{hh}}} = \frac{ M_{\rm Pl}^2 + \xi h^2}{M_{\rm Pl}} \frac{1}{\sqrt{M_{\rm Pl}^2 + \xi (1 + 6 \sigma \xi) h^2}} .
    \label{htochi}
\end{equation}
The rescaling in Eq.~(\ref{htochi}) depends on the underlying formulation of gravity. For the metric-gravity case ($\sigma = 1$), Eq.~(\ref{htochi}) may be solved analytically (see, e.g., Eq.~(3.7) in Ref.~\cite{Steingasser:2025txd}), though during inflation, for $h \gg M_{\rm Pl} / \sqrt{\xi}$ and $\xi \gg 1$, the solution simplifies to \cite{Bezrukov:2007ep}
\begin{equation}
    \chi (h) \simeq \sqrt{6} \, M_{\rm Pl} \, \ln \left( \frac{ \sqrt{\xi} \, h}{M_{\rm Pl}} \right) \quad {\rm (metric)} .
    \label{chimetric}
\end{equation}
Meanwhile, in the Palatini-gravity case ($\sigma = 0$), Eq.~(\ref{htochi}) may be solved analytically in a simple form for arbitrary $\xi$ (see Eq.~(3.16) in Ref.~\cite{Steingasser:2025txd}):
\begin{equation}
    \chi (h) = \frac{ M_{\rm Pl}}{\sqrt{\xi}} {\rm sinh}^{-1} \left( \frac{ \sqrt{\xi} \, h}{M_{\rm Pl}} \right) \quad {\rm (Palatini)}.
    \label{chiPalatini}
\end{equation}
See Fig.~\ref{fig:HiggsChiMaps} for plots of the mappings between $h$ and $\chi (h)$ for both metric and Palatini cases.

In both formulations (metric or Palatini), varying the effective action in the Einstein frame with respect to $g_{\mu\nu}$ and $\chi$ yields the coupled equations of motion,
\begin{equation}
    \ddot{\chi} + 3 H \dot{\chi} + V_E^\prime = 0
    \label{chieom}
\end{equation}
and
\begin{align}
    H^2 &= \frac{1}{ 3 M_{\rm Pl}^2} \left[ \frac{1}{2} \dot{\chi}^2 + V_E (\chi) \right] ,\\
    \dot{H} &= - \frac{1}{ 2 M_{\rm Pl}^2} \dot{\chi}^2 ,
\end{align}
where a prime denotes $d / d\chi$ and an overdot denotes $d / dt$, and $t$ is cosmic time. Here $H (t) \equiv \dot{a} / a$ is the Hubble parameter and $a(t)$ is the scale factor of the (spatially flat) Friedmann-Lema\^{i}tre-Robertson-Walker (FLRW) background metric. 

\begin{figure}
    \centering
    \includegraphics[width=0.95\linewidth]{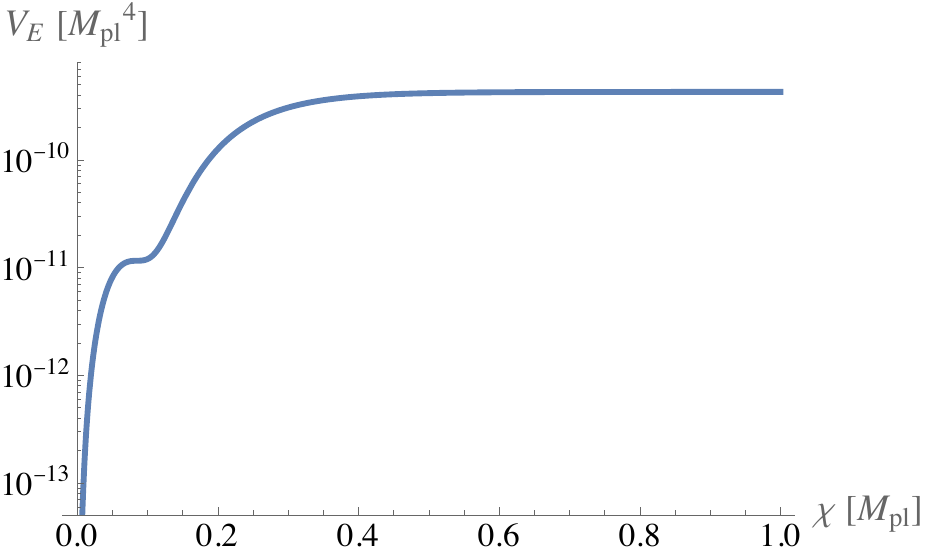}
    \caption{\justifying The effective potential in the Einstein frame, $V_E (\chi)$, for the case of Palatini gravity, with $\{\lambda_{\rm ll} , \delta, \xi, \mu_{\rm m} \} = \{10^{-5} , 10^{-2} , 40, 10^{-1} \, M_{\rm Pl} \}$.}
    \label{fig:VEPalatini}
\end{figure}

Inflation corresponds to a phase of accelerated expansion, $\ddot{a} > 0$. This can occur when the potential energy dominates the field's kinetic energy. The potential in near-critical Higgs inflation easily yields such dynamics, given the gentle plateau of $V_E (\chi)$ at large field values, as well as the near-critical point at smaller field values. See Fig.~\ref{fig:VEPalatini}. We may further characterize the dynamics in terms of the slow-roll parameters
\begin{equation}
        \epsilon \equiv - \frac{ \dot{H}}{H^2} , \quad
        \eta \equiv 2 \epsilon - \frac{\dot{\epsilon}}{2 H \epsilon}  .
    \label{SRparamsdef}
\end{equation}
To address the usual shortcomings of the original big bang model, inflation must persist for at least $N_{\rm total} \simeq 60$ e-folds, where $N (t) \equiv \int dt \, H$. Inflation ends when $\epsilon (t_{\rm end}) = 1$, which corresponds to $\ddot{a} (t_{\rm end}) = 0$. (See, e.g., Refs.~\cite{Bassett:2005xm,Lyth:2009imm,Guth:2013sya,Martin:2013tda,Baumann:2022mni,Kinney:2026qtr}.)

Fluctuations in the field $\chi$ source curvature perturbations, whose evolution we may analyze in terms of the quantity 
\begin{equation}
{\cal R} \equiv \psi + \frac{ H}{\dot{\chi}} \delta \chi .
\label{Rdef}
\end{equation}
Here $\psi$ is the comoving scalar metric perturbation on equal-time hypersurfaces. Both $\psi$ and $\delta \chi$ are gauge-dependent; the combination ${\cal R}$ is gauge-invariant to linear order in fluctuations. (See, e.g., Ref.~\cite{Kaiser:1994vs,Bassett:2005xm,Lyth:2009imm,Baumann:2022mni,Kinney:2026qtr}.) In the absence of any turning among the background fields, the isocurvature perturbations ${\cal S}^I$ (which are proportional to the fluctuations $\delta \phi^I$ for $I \geq 1$) decouple \cite{Kaiser:2012ak}, and modes of comoving wavenumber $k$ of the gauge-invariant curvature perturbation evolve as
\begin{equation}
    \ddot{\cal R}_k + \left( 3 + 4 \epsilon - 2 \eta \right) H \dot{\cal R}_k + \frac{ k^2}{a^2} {\cal R}_k = 0 .
    \label{Reom}
\end{equation}
Whenever $\eta < \frac{3}{2} + 2 \epsilon$, the modes ${\cal R}_k$ are damped by Hubble friction and their amplitude effectively freezes once they cross outside the Hubble radius during inflation, at the time $t_*$ such that $k_* = a(t_*) H(t_*)$. We consider predictions for various observables on scales near the CMB pivot scale, $k_* = 0.05 \, {\rm Mpc}^{-1}$. 

To compare with high-precision measurements of CMB anisotropies, we consider the dimensionless power spectrum. Within the single-field attractor, this may be written as
\begin{equation}
{\cal P}_{\cal R} (k, t_{\rm end}) = \frac{ k^3}{2 \pi^2} \left\vert {\cal R}_k (t_{\rm end}) \right\vert^2 .
\label{PRdef}
\end{equation}
For ordinary slow-roll evolution, we may consider the amplitude of the primordial power spectrum near the CMB pivot scale $k_*$,
\begin{equation}
A_s (k_*) \equiv {\cal P}_{\cal R} (k_*, t_{\rm end}) \simeq \frac{ H^2 (t_*)}{8 \pi M_{\rm Pl}^2 \epsilon (t_*)} ,
\label{Asdef}
\end{equation}
its spectral index $n_s$, 
\begin{equation}
    \begin{split}
        n_s (k_*) &\equiv 1 + \frac{ d \ln {\cal P}_{\cal R} (k, t_{\rm end})}{d \ln k} \bigg\vert_{k_*} \\
        &\simeq 1 - 6 \epsilon (t_*) + 2 \eta (t_*) ,
    \end{split}
\label{nsdef}
\end{equation}
the running of the spectral index $\alpha_s$, 
\begin{equation}
        \alpha_s (k_*) \equiv \frac{ dn_s}{d \ln k} \bigg\vert_{k_*}  \simeq \frac{ \dot{n}_s (t_*)}{H  (t_*)} , 
\label{alphadef}
\end{equation}
and the tensor-to-scalar ratio $r$,
\begin{equation}
        r \equiv \frac{ {\cal P}_T (k, t_{\rm end})}{ {\cal P}_{\cal R} (k, t_{\rm end} )} \bigg\vert_{k_*} \simeq 16 \epsilon (t_*),
\label{rdef}
\end{equation}
where ${\cal P}_T (k, t_{\rm end})$ is the dimensionless power spectrum of primordial tensor modes evaluated at the end of inflation. In Eqs.~(\ref{Asdef})--(\ref{rdef}), the expressions given in terms of the slow-roll parameters hold when the system undergoes ordinary slow-roll evolution, with $\vert \epsilon\vert , \vert \eta \vert \ll 1$ \cite{Bassett:2005xm}. Meanwhile, evolution within the single-field attractor suppresses typical multifield features, such as the fraction of power in isocurvature perturbations $\beta_{\rm iso} (k_*)$ and primordial non-Gaussianity $f_{\rm NL}$ \cite{Kaiser:2012ak,Kaiser:2013sna,Schutz:2013fua}.

Modes near $k_*$ cross outside the Hubble radius at the time \cite{Dodelson:2003vq,Liddle:2003as}
\begin{equation}
    N_* \simeq 62 + \frac{1}{4} \ln \left( \frac{ V_E^2 (t_*)}{3 M_{\rm Pl}^6 H^2 (t_{\rm end})} \right) - \frac{ N_{\rm reh}}{4} 
\label{NCMB}
\end{equation}
e-folds before the end of inflation. Here $N_{\rm reh}$ denotes the duration of reheating after the end of inflation~\cite{Amin:2014eta}. Models like Higgs inflation reheat efficiently, with $N_{\rm reh} \lesssim 3$ \cite{Bezrukov:2008ut,Garcia-Bellido:2008ycs,Sfakianakis:2018lzf,DeCross:2015uza,DeCross:2016fdz,DeCross:2016cbs,Nguyen:2019kbm,vandeVis:2020qcp,Bettoni:2021zhq,Laverda:2023uqv,Bettoni:2024ixe,Figueroa:2024asq,Chowdhury:2025mye}, so in this work we calculate predictions for CMB observables upon evaluating $N_*$ in Eq.~(\ref{NCMB}) for a given set of parameters (which affect $V_E (t_*)$ and $H (t_{\rm end})$), while neglecting $N_{\rm reh}$. Across most of parameter space of interest here, we find $N_* \simeq 56$ e-folds before the end of inflation. 

Predictions from models of inflation can be compared with high-precision measurements of CMB anisotropies, which yield \cite{Planck:2018jri,BICEP:2021xfz}
\begin{equation}
\begin{split}
A_s  &= \left( 2.09 \pm 0.10 \right) \times 10^{-9} , \\
n_s  &= 0.9649 \pm 0.0042 , \\
\alpha_s &= -0.0045 \pm 0.0067 , \\
r &< 0.036 ,
\end{split}
\label{CMBobservables}
\end{equation}

each evaluated at the CMB pivot scale $k_* = 0.05 \, {\rm Mpc}^{-1}$. Here we use the best-fit values from the {\it Planck} 2018 analysis, which remain consistent with the more recent reports from the ACT \cite{AtacamaCosmologyTelescope:2025nti} and SPT \cite{SPT-3G:2025bzu} collaborations \cite{Ferreira:2025lrd}. The small shifts in central values for quantities such as $n_s$ among Refs.~\cite{Planck:2018jri,AtacamaCosmologyTelescope:2025nti,SPT-3G:2025bzu} do not have any impact on our overall conclusions regarding critical Higgs inflation. When matching 

Finally, we note that in models that include a near-inflection point, such as critical Higgs inflation, the system may depart from ordinary slow-roll evolution at some time before inflation ends. In particular, in the vicinity of the region in which $V_E^\prime (\chi) \simeq 0$, the parameter $\epsilon \rightarrow 0^+$ and $\eta \rightarrow 3$, giving rise a phase known as \textit{ultra-slow-roll} evolution \cite{Kinney:2005vj,Martin:2012pe,Dimopoulos:2017ged,Pattison:2018bct}. Given the form of the equation of motion for modes ${\cal R}_k$ in Eq.~(\ref{Reom}), we see that modes of comoving $k$ for which $k \ll aH$ while $\eta > \frac{3}{2} + 2 \epsilon$, during ultra-slow-roll, will be amplified. This can yield observable signatures such as enhanced gravitational waves---induced at second-order in perturbation theory by the large amplitude of ${\cal R}_k$---and the production of primordial black holes soon after the end of inflation. (For reviews, see Refs.~\cite{Allahverdi:2020bys,Domenech:2021ztg,Escriva:2022duf,Ozsoy:2023ryl}.) Previous authors have indeed identified regions of parameter space in models of critical Higgs inflation in which each of these phenomena can occur \cite{Ezquiaga:2017fvi,Bezrukov:2017dyv}. 

Given our interest in this article in the three criteria---consistency with precision EW-scale measurements, CMB observables, and limited sensitivity to unknown UV physics---it suffices in our case to consider the threshold correction factor $\delta \sim {\cal O} (10^{-2})$. In our analysis, therefore, we never consider points in parameter space that get arbitrarily close to the actual inflection point ($\delta = 0$). Many previous studies have indicated that departures from ordinary slow-roll evolution in (effectively) single-field inflation, of the sort that can source detectable gravitational waves and/or primordial black holes, require fine-tuning at least one dimensionless parameter to one part in ${\cal O} (10^{4})$ or below. (See, e.g., Refs.~\cite{Ezquiaga:2017fvi,Bezrukov:2017dyv,Garcia-Bellido:2017mdw,Geller:2022nkr,Qin:2023lgo,Cole:2023wyx,Lorenzoni:2025gni,Lorenzoni:2024krn}.) In the present case, by limiting attention to $\delta \sim {\cal O} (10^{-2})$, we confirm that the system never leaves ordinary slow-roll, and hence the regions of parameter space on which we focus produce neither detectable gravitational waves nor primordial black holes.

\section{Perturbative Unitarity and sensitivity to new physics}
\label{sec:Unitarity}

One of the biggest conceptual appeals of Higgs inflation is the possibility to probe the inflaton field at relatively low energies. A necessary condition for establishing this connection is that the \emph{same} underlying model serves as a reliable effective description across vastly different regimes---in particular, near both $\langle h \rangle=v_{\rm EW} \sim {\cal O} (10^2 \, {\rm GeV}) \ll M_{\rm Pl}$ and $\langle h \rangle \sim h_{\rm infl}\sim M_{\rm Pl}/\sqrt{\xi}$. For ordinary, tree-level Higgs inflation, earlier investigations have shown that scattering of Higgs particles with energies larger than some cutoff $\Lambda$ would lead to a violation of perturbative unitarity \cite{Burgess:2009ea,Burgess:2010zq,Barbon:2009ya,Hertzberg:2010dc,Bezrukov:2010jz,Giudice:2010ka,Lerner:2011it,Bezrukov:2012hx,Bezrukov:2013fka,Calmet:2013hia,Barbon:2015fla,Steingasser:2025txd}. This issue arises from the relatively large values of $\xi$ necessary for predictions from tree-level Higgs inflation to match cosmological observables. The ability of near-critical Higgs inflation to yield the correct CMB observables for significantly smaller values of $\xi$ thus suggests that it might avoid these unitarity issues. In this section, we give a brief review of a recent high-precision unitarity analysis capable of capturing the near-critical regime \cite{Steingasser:2025txd}, before applying these results in Sec.~\ref{sec:NumericalResults}.

The violation of perturbative unitarity by the nonminimal coupling can best be understood by representing the coupling in terms of higher-dimensional operators following the conformal transformation of Eq.~\eqref{eq:conftransf}. For metric gravity with $\xi \gg 1$, and expanding around $\langle h \rangle \sim v_{\rm EW} \ll M_{\rm Pl} / \sqrt{\xi}$, the most important terms are the nondiagonal elements of the field-space metric ${\cal G}_{IJ} (\phi^K)$ in Eq.~\eqref{eq:GIJ},
\begin{equation}
\begin{split}
     \frac{\mathcal{L} }{\sqrt{-g}} \supset & \>3 \frac{\xi^2 M_{\rm Pl}^2}{(M_{\rm Pl}^2+ 2\xi |\mathbf{H}|^2 )^2}   \delta_{IK} \delta_{JL} \phi^K \phi^L \partial_\mu \phi^I \partial_\nu \phi^J \nonumber \\ 
     \simeq & \>3 \frac{\xi^2 }{M_{\rm Pl}^2}   \delta_{IK} \delta_{JL} \phi^K \phi^L \partial_\mu \phi^I \partial_\nu \phi^J , \label{eq:dLGIJ}
\end{split}
\end{equation}
where we have neglected terms suppressed by higher powers of $M_{\rm Pl}/\sqrt{\xi}$ in the last step. In this section we will briefly consider the dominant contribution to scattering matrix elements that arises from this particular interaction; see Ref.~\cite{Steingasser:2025txd} for a more complete discussion.

Eq.~\eqref{eq:dLGIJ} gives rise to a four-point interaction between different species of scalars within the Higgs doublet. Through dimensional analysis alone, it is easy to see that the corresponding S-matrix element for the scattering of two scalar particles with center-of-mass energy $E$ scales as
\begin{align}\label{eq:MSsim}
    \mathcal{M}_{2 \to 2} \sim \left( \frac{E\cdot \xi}{M_{\rm Pl}} \right)^2,
\end{align}
with each factor of $E$ arising from one of the derivatives in Eq.~\eqref{eq:dLGIJ}. Perturbative unitarity of the S-matrix requires that $|\mathcal{M}_{2 \to 2}|^2<1$. Restoring the prefactors dropped in Eq.~\eqref{eq:MSsim}, this implies that the Lagrangian can only consistently be valid for energies~\cite{Steingasser:2025txd}
\begin{align}\label{eq:EmaxMetric}
    E<E_{\rm max}^{\rm EW}\simeq 2.05 \, \frac{M_{\rm Pl}}{\xi} \quad\quad (\text{metric}).
\end{align}
Demanding perturbative unitarity therefore implies that the theory needs to be modified above some intermediate scale $\Lambda<E^{\rm EW}_{\rm max}$ in such a way that $\vert \mathcal{M}_{2 \to 2} \vert^2$ remains smaller than $1$. Yet any new physics, beyond the terms in Eqs.~(\ref{Vcritical1}), (\ref{kineticsector}), and (\ref{eq:GIJ}) that could alter these scattering amplitudes could \emph{also} affect the inflationary dynamics. This would render cosmological predictions from the model dependent upon the details of the UV completion~\cite{Barbon:2009ya}.

A similar argument applies in the case of Higgs inflation in the Palatini formulation of gravity. As noted below Eq.~(\ref{eq:GIJ}), the Palatini case corresponds to $\sigma = 0$, so that the term in Eq.~\eqref{eq:dLGIJ} remains absent. For small $\lambda_{\rm nl}$, the relevant scattering amplitudes in the Palatini case arise instead from terms suppressed by powers of $E/ (\sqrt{\xi}\, M_{\rm Pl})$. In Ref.~\cite{Steingasser:2025txd}, it was shown that this leads to an upper bound
\begin{align}\label{eq:EmaxPalatini}
    E<E_{\rm max}^{\rm EW}\simeq 4.94 \, \frac{M_{\rm Pl}}{\sqrt{\xi}} \quad\quad (\text{Palatini}).
\end{align}
Although the scaling of $E_{\rm max}^{\rm EW}$ with $\xi$ is parametrically weaker for the Palatini case compared to the metric-gravity case, we will see in Sec.~\ref{sec:NumericalResults} that successful inflation in the Palatini case typically requires much larger values of $\xi$. We must therefore consider perturbative unitarity carefully for each model.

Before applying these results to near-critical Higgs inflation, it is important to note two conceptual subtleties. First, the results in Eqs.~(\ref{eq:EmaxMetric}) and (\ref{eq:EmaxPalatini}) were obtained by considering the scattering of high-energy Higgs particles in a Minkowski background, in the vicinity of $\langle h \rangle \sim v_{\rm EW}$, rather than around the inflationary condensate $\langle h \rangle \sim h_{\rm infl} \sim M_{\rm Pl} /\sqrt{ \xi}$. The regime $\langle h \rangle \sim v_{\rm EW}$ is relevant to consider because a single model of Higgs inflation is supposed to provide a reliable effective description across an enormous range of scales, from $\langle h \rangle \sim v_{\rm EW}$ all the way up to $\langle h \rangle \sim h_{\rm infl}$. If one were to require instead \emph{only} that the model serve as a reliable effective description for dynamics during and immediately after cosmic inflation, for field values in the vicinity of $\langle h \rangle \sim h_{\rm infl}$, perturbative unitarity would require new physics to enter at some scale $\Lambda \leq E_{\rm max}^{\rm infl}$ with $E_{\rm max}^{\rm infl} \gg E_{\rm max}^{\rm EW}$, as indicated in Eq.~(\ref{Emaxinfl}). In such cases, (critical) Higgs inflation does indeed behave as a self-consistent effective theory. (See esp. Sec.~6 of Ref.~\cite{Steingasser:2025txd}.) 

Second, in the absence of a specific UV completion, it is unclear how the BSM terms would enter the effective action. The relevant terms could be represented by a tower of higher-dimensional operators involving the magnitude of the Higgs doublet $\vert {\bf H} \vert \sim h$, or the canonically normalized field $\chi (h)$, or even---given the form of the effective potential in Eq.~(\ref{eq:VE})---the term 
\begin{equation}
F (h) \equiv \frac{ h }{ \sqrt{ 1 + \xi h^2 / M_{\rm Pl}^2} } = z (h) \cdot \mu_{\rm m},
\label{Fdef}
\end{equation}
where the dimensionless quantity $z (h)$ is defined in Eq.~(\ref{eq:zdef}).\footnote{Any attempt to describe the UV-complete theory in terms of $F (h)$ rather than $h$ or $\chi$ would be difficult, given that the kinetic term cannot be expressed in terms of this auxiliary field.} 

We therefore consider three proxy measures with which to assess the degree to which predictions for cosmological observables are likely to depend on the (necessary but unknown) UV physics. We optimistically set $\Lambda \simeq E_{\rm max}^{\rm EW}$, saturating the more general requirement $\Lambda \leq E_{\rm max}^{\rm EW}$. Then we evaluate the ratios of various measures of the field around the time when CMB-relevant perturbations cross outside the Hubble radius during inflation, at time $t_*$, with $\Lambda \simeq E_{\rm max}^{\rm EW}$:
\begin{equation}
    \frac{ h (t_* )}{E_{\rm max}^{\rm EW}} \> , \>\> \frac{ \chi (t_*)}{E_{\rm max}^{\rm EW}} \>, \>\> \frac{ F (t_*)}{E_{\rm max}^{\rm EW}} .
    \label{UVproxies}
\end{equation}
A minimal requirement for cosmological predictions to remain insensitive to the unknown UV physics is for any of these ratios to remain less than 1.

Such a requirement is a fairly weak proxy for the likely sensitivity to UV physics. Consider, as an example, a typical next-to-leading-order contribution to the effective potential, $C_H \vert {\bf H}\vert^6 / \Lambda^2$. If $\Lambda \gg \vert {\bf H} \vert \sim h$ during inflation, then the effect of this operator will be suppressed by the small ratio of these scales. Yet requiring $h (t_*) \ll \Lambda \simeq E_{\rm max}^{\rm EW}$ may not be sufficient to render predictions for cosmological observables insensitive to this new operator. In particular, the ratio of this new term $\Delta V_{\rm UV}$ to the critical Higgs potential $V_E (h)$ takes the form
\begin{equation}\label{eq:sensest}
    \frac{ C_H h^6 / \Lambda^2}{\lambda_{\rm eff} h^4 / 4} \simeq \frac{ 4 C_H}{\lambda_{\rm eff}} \frac{ h^2}{(E_{\rm max}^{\rm EW})^2} \sim 3 \cdot 10^6 \,C_H \frac{ h^2}{(E_{\rm max}^{\rm EW} )^2} ,
\end{equation}
where the last step uses $\lambda_{\rm eff} \simeq \lambda_{\rm ll} / 16$ near the critical point, along with the central value of $\lambda_{\rm ll} (\mu_{\rm m})$ reported in Table~\ref{table:confidence}. Even if $h (t_*) \ll E_{\rm max}^{\rm EW}$, contributions from the UV completion could be comparable to (or considerably greater than) the SM terms unless the Wilson coefficient were exponentially small, $C_H \ll {\cal O} (10^{-6})$. This highlights an important trade-off when comparing the unitarity properties of near-critical and tree-level Higgs inflation. Whereas the former may reproduce the observed CMB observables with a smaller value of nonminimal coupling $\xi$ (and hence a larger value of $E_{\rm max}^{\rm EW}$), this requires a two-loop-suppressed effective quartic coupling $\lambda_{\rm eff}$, thus making the potential more sensitive to contributions from the UV completion.

Nonetheless, in the absence of a specific UV completion that would preserve perturbative unitarity at high energies, we consider the three proxies identified in Eq.~(\ref{UVproxies}). In Sec.~\ref{sec:NumericalResults} we consider each of these proxies for both the metric and Palatini formulations of critical Higgs inflation.

\section{Numerical Analysis}
\label{sec:NumericalResults}

\subsection{Numerical Methods}
\label{sec:NumericalMethods}
Our goal is to discover whether any part of the parameter space for (near-)critical Higgs inflation can \textit{simultaneously} satisfy all three of the criteria described in Sec.~\ref{sec:EvaluationCriteria}.  We treat all points within the full SM 3$\sigma$-compatible region identified in Sec.~\ref{sec:HiggsEWscale} as equally consistent with EW-scale measurements; that is, we do not weight central values over peripheral ones. For each parameter tuple $\{\lambda_{\rm ll}, \mu_{\rm m}, \delta \}$ within the SM 3$\sigma$ region, we find the value of $\xi_*$ that reproduces the central value for ${\cal P}_{\cal R}(k_*)=A_s$ while leaving $n_s, \alpha_s, r$ within $2\sigma$ of their observed values, as reported in Eq.~\eqref{CMBobservables}. 

We proceed in two stages: first, we evaluate $A_s(\xi)$ on a coarse logarithmic grid by solving the background field equations at each point on the grid. We  identify all sign changes of the residual $r=A_s(\xi)-A_s^{\rm obs}$ and bracket each sign change, then refine each bracketed region to home in on the value of $\xi_*$ at which $r=0$ using Brent's method. This procedure is described in considerably more detail in Appendix~\ref{sec:app_rootfinding}. (The residual $r$ should not be confused with the tensor-to-scalar ratio $r (k_*)$.) There will be at least one and at most three such roots for any given tuple $\{ \lambda_{\rm ll},
\mu_{\rm m},\delta \}$. We find that typically at least one root yields predictions for all CMB observables that fall within $2\sigma$ of the best-fit values in Eq.~(\ref{CMBobservables}). In every case, we select the smallest CMB-compatible root $\xi_*$, so as to yield the largest allowable value of $\Lambda = E_{\rm max}^{\rm EW}$, thereby minimizing the likely dependence of model predictions on unknown UV physics, as parameterized by the quantities $h_* / \Lambda, \chi_* / \Lambda$, and $F (h_*) / \Lambda$ of Eq.~(\ref{UVproxies}).

\subsection{Shape of the Potential in Metric Gravity}\label{subsec:shapeofpotential}

First we consider the dependence of the inflationary dynamics on the parameters that control the shape and normalization of the near-critical potential in the metric gravity formulation. For fixed $\mu_{\rm m}$, $\delta$, and $\xi$, the parameter $\lambda_{\rm ll}$ corresponds to a rescaling of the overall potential,
\begin{figure*}[t!]
    \centering
    \includegraphics[width=\linewidth]{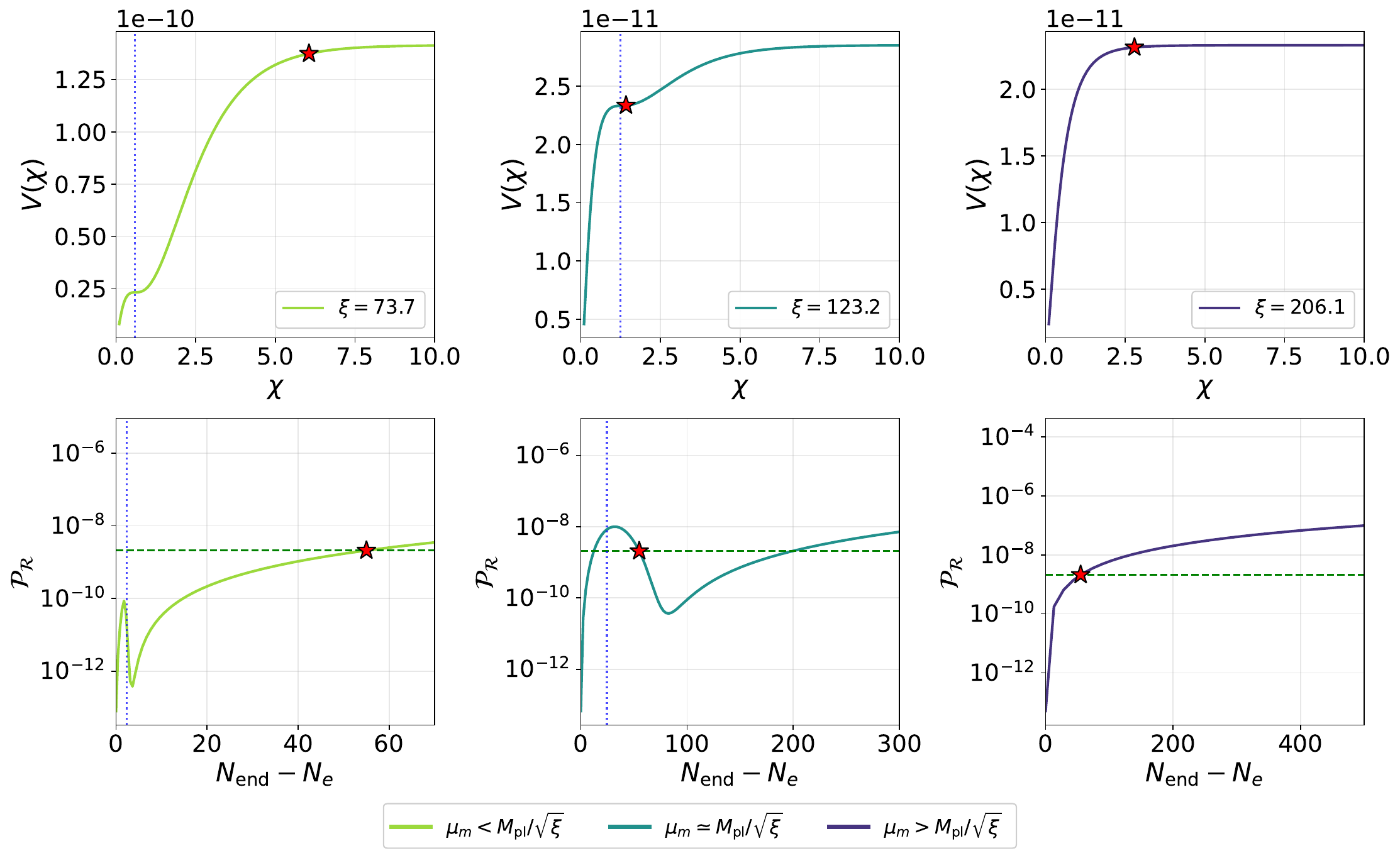}
    \caption{\justifying \emph{(Top row)} The effective potential in the Einstein frame $V_E$ as a function of the canonically normalized field $\chi$ for the metric-gravity case, in units of $M_{\rm Pl}$. Red stars indicate the field value at the time $t_*$ when CMB-relevant perturbations exit the Hubble radius.
     Vertical dotted blue lines indicate the location of the inflection point, $V_E^{\prime}(\chi)=V_E^{\prime\prime}(\chi)=0$. From left to right, the potential is shown for $\xi =73.7, 123.2,$ and $206.1$ at fixed $\lambda_{\rm ll}=2.7\times10^{-5},\, \mu_{\rm m}= 0.093\, M_{\rm Pl}, \,\delta = 0.01$. \emph{(Bottom row)} Dimensionless power spectrum of the gauge-invariant curvature perturbations ${\cal P}_{\cal R}$ as a function of e-folds after the start of inflation. The green dashed horizontal line is the observed amplitude $A_s = 2.09\times10^{-9}$; red stars mark the number of e-folds before the end of inflation when the CMB pivot scale exits the Hubble radius; and the blue dotted lines show the number of e-folds before the end of inflation at which the inflaton encounters the inflection point.\textit{(Left to right)} For $\mu_{\rm m} < M_{\rm Pl}/\sqrt{\xi}$ (\emph{light green}), the inflection point sits at field values $\chi \ll \chi_*$, toward the end of inflation, and generates a feature in the power spectrum at length scales exponentially shorter than the CMB pivot scale, leaving CMB observables intact. When $\mu_{\rm m}\simeq M_{\rm Pl}/\sqrt{\xi}$ (\emph{blue}), the inflection point occurs at or very near the field values at which the CMB pivot scale exits the Hubble radius, such that the resultant peak in ${\cal P}_{\cal R}$ occurs near the CMB scale, which typically pushes the spectral index $n_s$ out of compliance with observations. In such cases, however, one may still identify points in parameter space for which predictions for all CMB observables remain consistent (within $2 \sigma$) of the best-fit values in Eq.~(\ref{CMBobservables}), using the techniques described in Appendix~\ref{sec:app_rootfinding}. For $\mu_{\rm m} > M_{\rm Pl}/\sqrt{\xi}$ (\emph{purple}), the inflection point occurs at $\chi\gg\chi_*$ (not visible) and has no effect on the dynamics within the observable window of inflation.
  }
    \label{fig:6panel}
\end{figure*}

\begin{equation}
V_E(h)=\frac{\lambda_{\rm ll}\,\mu_{\rm m}^4}{4}
\left[\frac{1+\delta}{16}+\ln^2 z(h)\right]z^4(h).
\end{equation}
To first order in the slow-roll parameters, we may substitute $3 M_{\rm Pl}^2 H^2 \simeq V_E (\chi)$ and $2\epsilon \simeq M_{\rm Pl}^2(V_E^\prime / V_E)^2$ (where the prime indicates a derivative with respect to the canonically normalized field $\chi$), so that Eq.~(\ref{Asdef}) may be written as
\begin{equation}
{\cal P}_{\cal R}(k_*) = A_s (k_*) \simeq
\frac{V_E(\chi_*)}{24\pi^2M_{\rm Pl}^4\epsilon(\chi_*)}
\propto \lambda_{\rm ll}.
\label{PRSRV}
\end{equation}
We consistently find such a linear scaling, $A_s \propto \lambda_{\rm ll}$, throughout parameter space. 

In contrast, the shape of $V_E (h)$, and hence the predictions for CMB observables, varies nonlinearly with the scale $\mu_{\rm m}$. In the critical limit $\delta=0$, the location of the inflection point is given by $z_{\rm crit}=e^{-1/4}$. 
Inverting Eq.~\eqref{eq:zdef} gives
\begin{align}
    h_{\rm crit} &=  \frac{\mu_{\rm m} \,z_{\rm crit}}{\sqrt{1- z_{\rm crit}^2\xi \frac{\mu_{\rm m}^2}{M_{\rm Pl}^2}}}  \nonumber \\ 
    & \begin{cases}
         \simeq \mu_{\rm m} \,e^{-1/4} \quad {\rm for} \quad  \mu_{\rm m} \ll M_{\rm Pl}/\sqrt{\xi}\\
        \to \infty \quad {\rm for} \quad \mu_{\rm m} \to e^{1/4} M_{\rm Pl}/\sqrt{\xi} 
    \end{cases}\, ,
\end{align}
where the latter limit corresponds to the upper limit in Eq.~(\ref{inflationpointbound}). 
The dependence of $h_{\rm crit}$ on $\mu_{\rm m}$ effectively divides the parameter space in the metric formulation into three dynamical regimes, illustrated in Fig.~\ref{fig:6panel}. These three regimes are:
\begin{enumerate}
    \item[(i)] $\mu_{\rm m} \ll M_{\rm Pl}/\sqrt{\xi}$: the inflection point occurs at small values of the canonical field $\chi$, and occurs toward the end of inflation, thus is distant from the large-field plateau and from the field value $\chi_*$ at which the CMB pivot scale exits the Hubble radius. In this regime, the dynamics near the inflection point only matter within the final few e-folds of inflation and only amplify the power spectrum of primordial curvature perturbations on length scales exponentially shorter than the CMB pivot scale. All CMB observables remain well within $2\sigma$ of their observed central values. 
    \item[(ii)] $\mu_{\rm m}\sim M_{\rm Pl}/\sqrt{\xi}$: the inflection point occurs near the field value $\chi_*$ at which CMB-relevant perturbations exit the Hubble radius. In this regime, the resulting feature in $\mathcal{P}_{\mathcal R}(k)$ can enter the observable window, giving rise to a nonmonotonic dependence of $A_s(k_*)$ on $\xi$. For a fixed combination $\{\lambda_{\rm ll},\mu_{\rm m},\delta\}$, the condition $A_s(k_*)=A_s^{\rm obs}$ may therefore admit three distinct roots in $\xi$. The middle root typically places the feature sufficiently close to the CMB pivot scale to drive $n_s$, and in some cases $r$, out of compliance with observational constraints. The remaining roots correspond to trajectories for which the feature is displaced from the CMB pivot scale; at least one of these roots yields predictions for the full set of CMB observables that remain within the observational bounds. We identify all such roots using the procedure described in Appendix~\ref{sec:app_rootfinding} and retain the smallest CMB-compatible value of $\xi$.
    \item[(iii)] $\mu_{\rm m} \gg M_{\rm Pl}/\sqrt{\xi}$: the inflection point occurs at field values $\chi \gg \chi_*$, and does not affect predictions for CMB observables.
\end{enumerate}

The fact that the shape of the potential---and hence the dynamics during inflation---depends on the scale $\mu_{\rm m}$ in complicated ways means that the relationship between model parameters and predictions for CMB observables is more complicated in critical Higgs inflation than in ordinary (tree-level) Higgs inflation. In the presence of loop corrections and a possible inflection point in the potential, values of parameters $\{ \lambda_{\rm nl}, \xi \}$ that would have yielded a close match to CMB observables in the tree-level case do not necessarily provide compliance with CMB measurements in the loop-corrected case. 

As we demonstrate in Secs.~\ref{sub:metric_results} and \ref{sub:Palatini_results}, in both the metric-gravity and Palatini-gravity formulations, one may adjust model parameters for critical Higgs inflation to satisfy {\it at most} two of three criteria: compliance with EW-scale measurements, consistency with CMB observations, and consistency with various unitarity proxies. In no cases do there exist points in parameter space for which critical Higgs inflation can satisfy all three criteria.

\subsection{Results in Metric Gravity}\label{sub:metric_results}

\begin{figure}
    \centering
    \includegraphics[width=\linewidth]{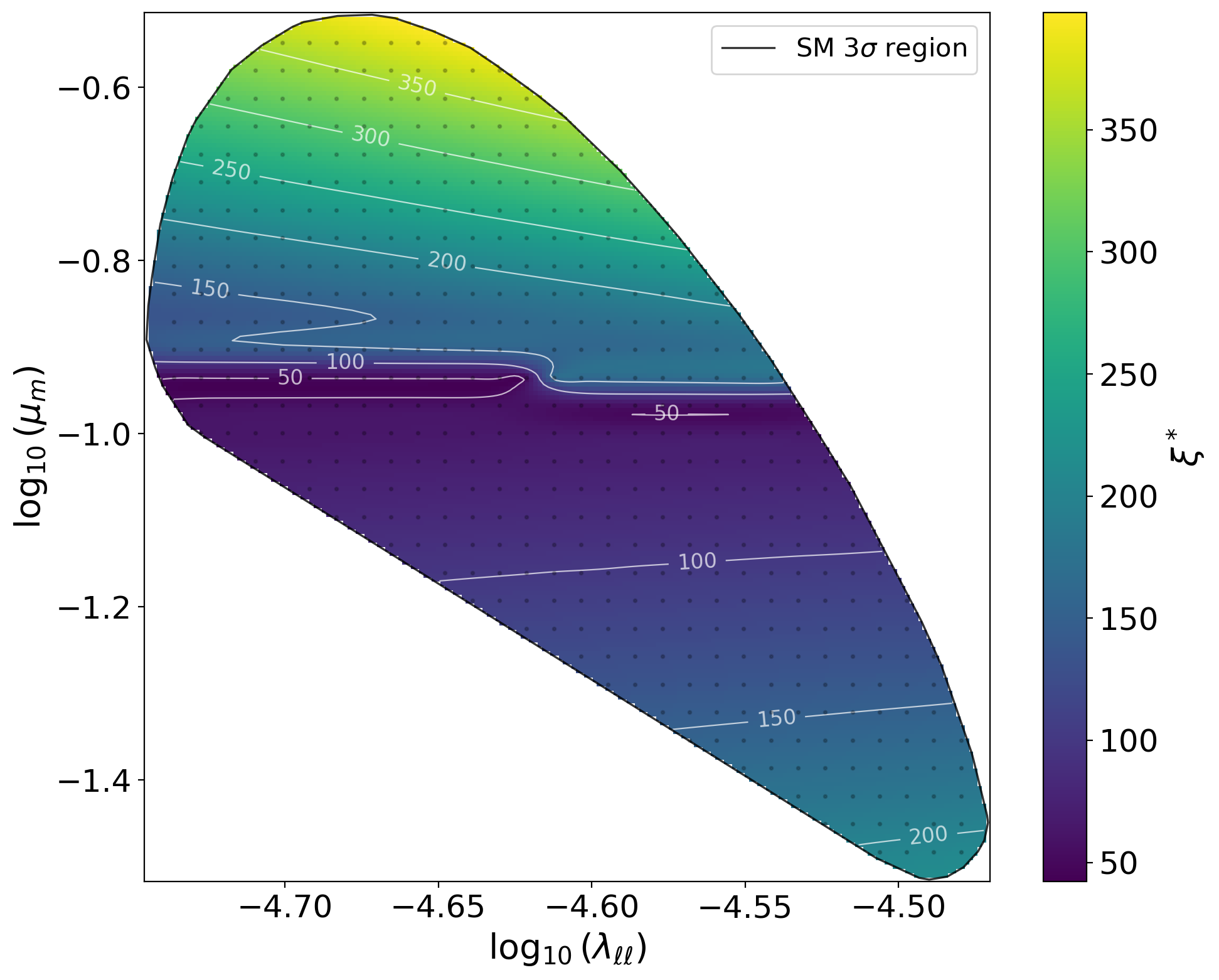}
    \caption{ \justifying
    Nonminimal couplings $\xi_*$ across the SM 3$\sigma$ region of $\{\lambda_{\rm ll},\mu_{\rm m}\}$, with $\delta= 0.01$ fixed, for critical Higgs inflation in the metric-gravity formulation. At each point in parameter space, $\xi_*$ is chosen in order to yield the observed amplitude of scalar curvature perturbations at CMB scales, $A_s = 2.09 \times 10^{-9}$. After fixing $\xi_*$, predictions for all other relevant CMB observables also remain consistent (within $2\sigma$) with the best-fit values in Eq.~(\ref{CMBobservables}), upon following the routine described in Appendix~\ref{sec:app_rootfinding}. Black dots show the same underlying grid as in Fig.~\ref{fig:chickenpoxplot}, and white contours outline $\xi_*$ values at integer multiples of 50, as in a topographical map. Across the entire $3\sigma$ SM region, we find $50 \lesssim \xi_* \lesssim 400$, with the smallest values attained in the region for which $\mu_{\rm m} \ll M_{\rm Pl}/\sqrt{\xi_*}$, corresponding to the ``smallest-root'' scenario described in 
    Appendix~\ref{sec:app_rootfinding}. The corresponding UV sensitivity proxies for these simultaneously SM- and CMB-compliant points are shown in Fig.~\ref{fig:proxiesdist}. 
    }
    \label{fig:xistarcompliance}
\end{figure}

\begin{figure} 
\centering \includegraphics[width=\linewidth]{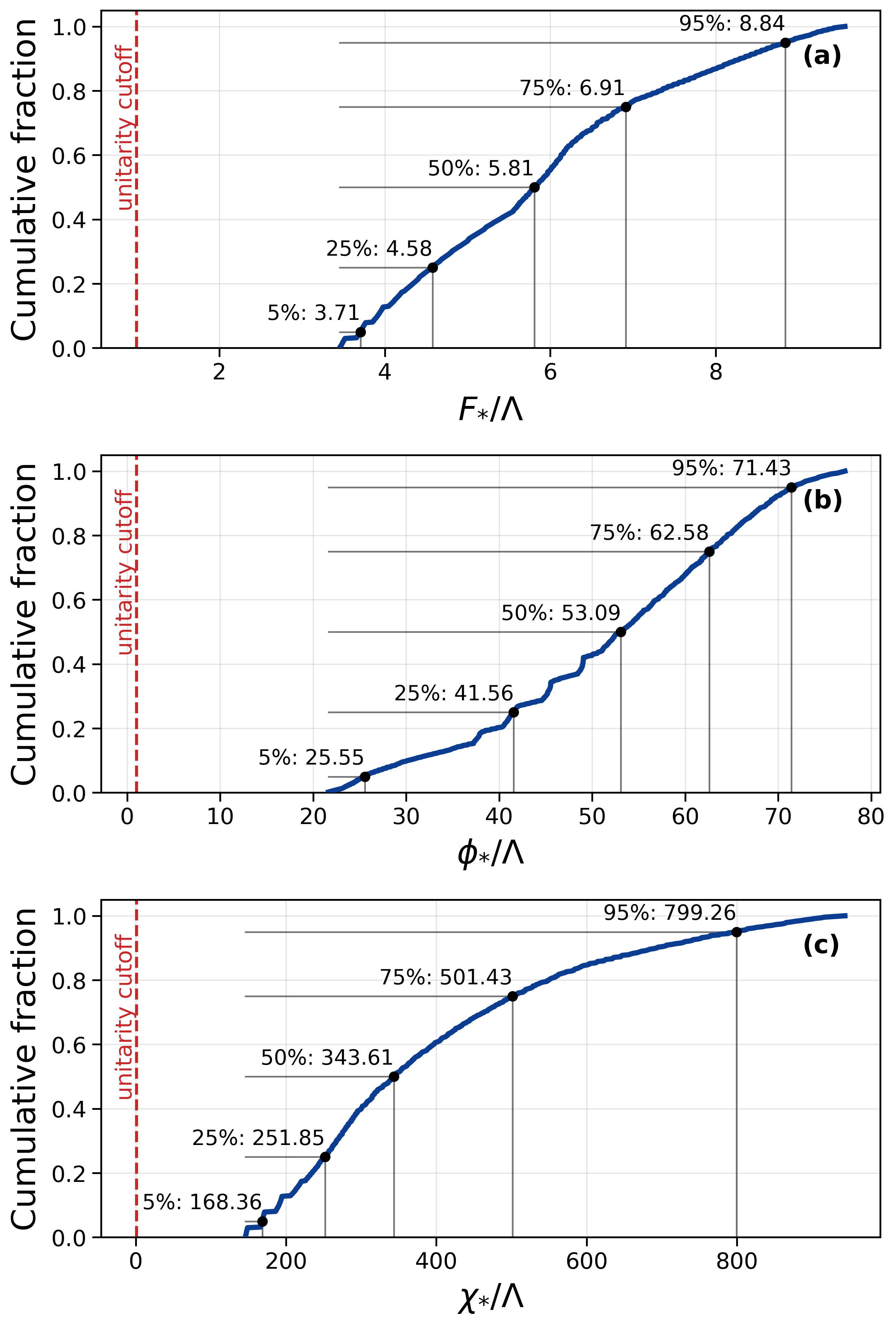}
    \caption{\justifying Cumulative distributions of the three proxies for sensitivity to unknown UV physics for the metric-gravity case, evaluated at the CMB pivot scale for the metric-gravity parameter points that are consistent with the SM $3\sigma$ region and with CMB observations. From top to bottom, the panels show $F_*/\Lambda$, $h_*/\Lambda$, and $\chi_*/\Lambda$, where $\Lambda = E_{\max}^{\rm EW} = 2.05 M_{\rm Pl}/\xi_*$. The red dashed line marks unity, corresponding to the perturbative-unitarity cutoff, while the labeled black markers indicate the 5th, 25th, 50th, 75th, and 95th percentiles. All three proxies exceed unity throughout the sampled region, indicating that predictions from the model for points in parameter space that are consistent with the SM $3\sigma$ requirements as well as with CMB observations are nonetheless generically sensitive to whatever (unknown) UV physics is required to restore unitarity.
 }
    \label{fig:proxiesdist}
\end{figure}

\begin{figure*}        
\centering                          
\includegraphics[width=0.48\linewidth]{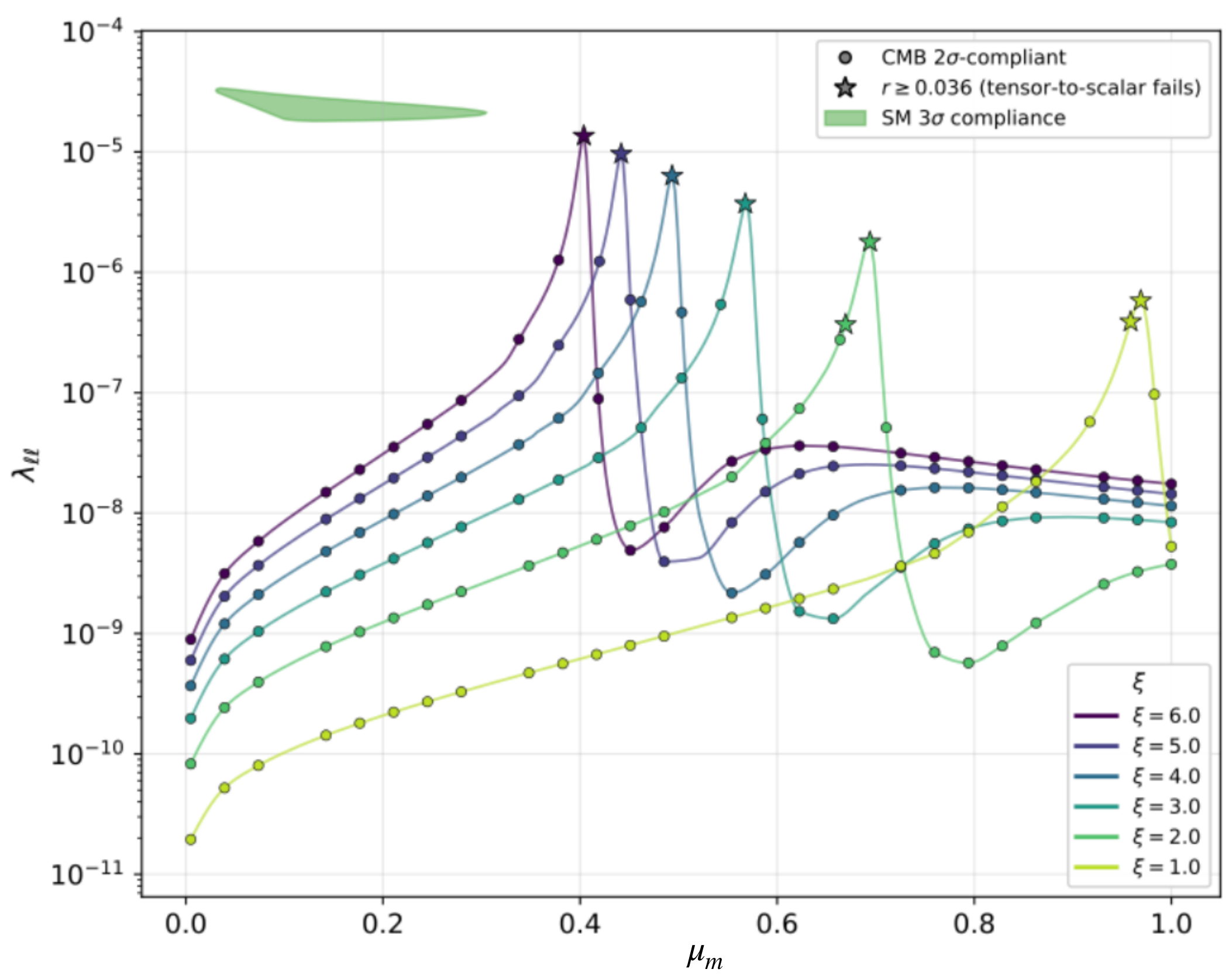}
    \includegraphics[width=0.48\linewidth]{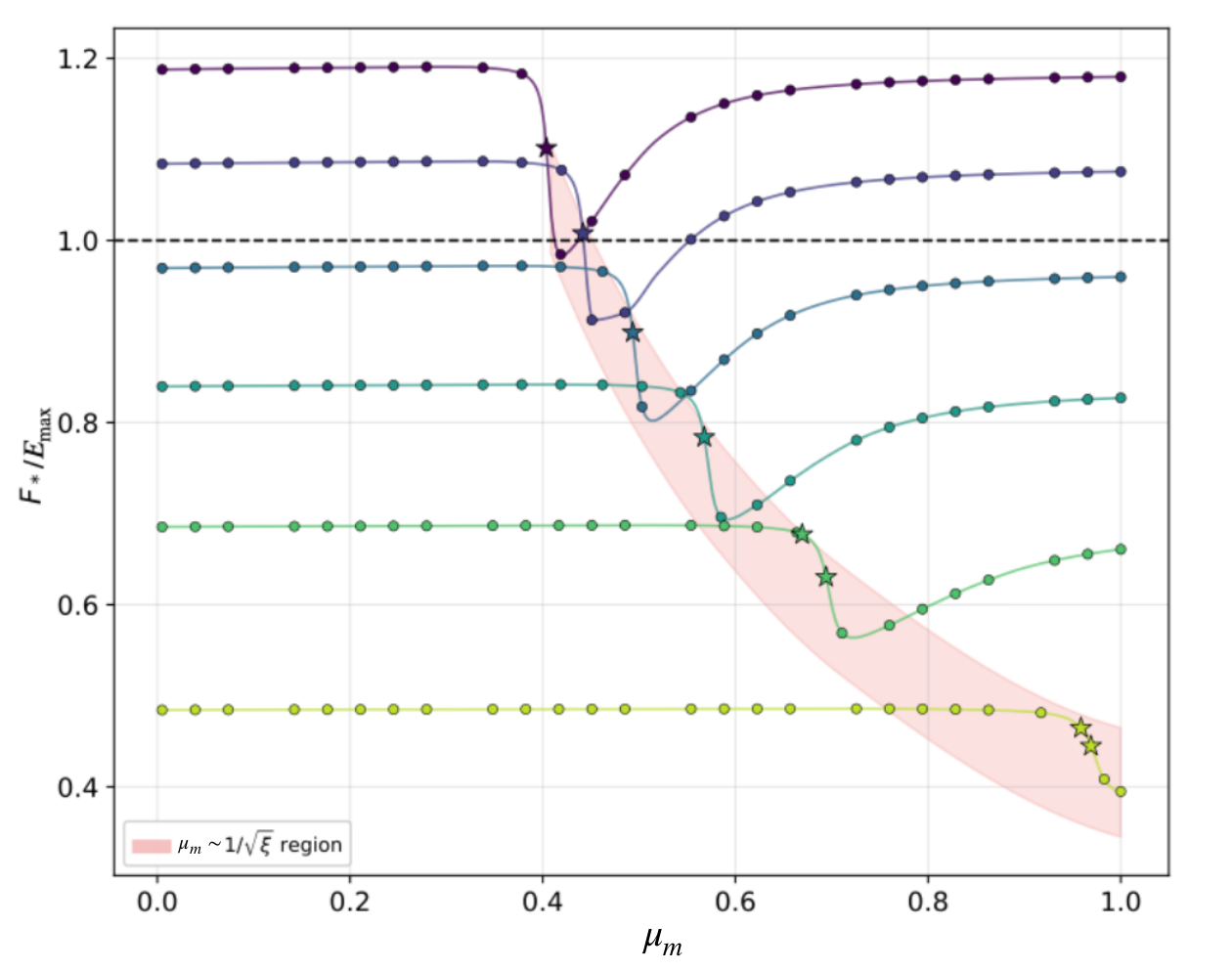}
    \caption{\justifying ({\it Left}) Correlation between the parameters $\mu_{\rm m}$ and $\lambda_{\rm ll}$, showing curves with fixed $\{\delta, \xi \}$ which maintain constant values for all CMB observables at each point for near-critical Higgs inflation in metric gravity. Circles indicate points in parameter space for which predictions for all CMB observables $\{A_s, n_s, \alpha_s, r \}$ fall within $2\sigma$ of current measurements. The curves demonstrate that a change in $\lambda_{\rm ll}$ can be compensated by a corresponding change in $\mu_{\rm m}$ to keep the predictions for CMB observables fixed. This relationship fails in one scenario: when the value of $\mu_{\rm m}$ becomes of order $M_{\rm Pl}/\sqrt{\xi_*}$ for a given curve, the curve spikes sharply. Stars indicate that points at or near the peaks of each curve exceed the observational bound on the tensor-to-scalar, yielding $r (k_*) > 0.036$. Note that \emph{none} of the curves shown here (with $\xi \leq 6$) lie within the SM $3 \sigma$ region ({\it solid green}). ({\it Right}) The UV sensitivity proxy $F_* / \Lambda$ for the same parameters shown on the left. The right-hand plot demonstrates that there are no CMB-compliant parameter sets within the SM $3\sigma$ region for which the UV sensitivity proxy $F_{*}/ \Lambda <1$. The largest value of $\xi$ for which $F_*/\Lambda <1$ at \emph{all values} of $\{\mu_{\rm m}, \lambda_{\rm ll} \}$ is $\xi \simeq 4.15$, and the largest value of $\xi$ for which \emph{at least one pair of values} $\{\mu_{\rm m},\lambda_{\rm ll}\}$ yields $F_*/\Lambda <1$ is $\xi \simeq 6.0$. As is evident from Fig.~\ref{fig:xistarcompliance}, the minimum value of $\xi$ compatible with constraints from both EW-scale measurements and CMB observables is $\xi \geq 50$, for which $F_* / \Lambda \gg 1$.}
    \label{fig:proxiesdist2}
\end{figure*}

We first consider regions of parameter space that are consistent with EW-scale measurements and that yield predictions for CMB observables consistent with recent measurements. In Fig.~\ref{fig:xistarcompliance} we fix $\delta = 0.01$ and restrict attention to values of $\{ \lambda_{\rm ll} , \mu_{\rm m} \}$ that lie within the SM $3\sigma$ region. (We consider other values of $\delta$ in Fig.~\ref{fig:roots_for_three_deltas}, and find qualitative agreement.) Then we identify values of the nonminimal coupling $\xi$ that yield the observed amplitude of scalar curvature perturbations $A_s (k_*)$. Across the entire SM $3 \sigma$ region, we find $50 \lesssim \xi \lesssim 400$. Upon fixing $\xi$ by matching $A_s (k_*)$, every point shown in Fig.~\ref{fig:xistarcompliance} yields predictions for \emph{all} relevant CMB observables within $2\sigma$ of the best-fit values given in Eq.~(\ref{CMBobservables}). (For those cases in which $\mu_{\rm m} \sim M_{\rm Pl} / \sqrt{\xi}$, we use the technique described in detail in Appendix~\ref{sec:app_rootfinding} to identify the smallest value of $\xi$ among three roots of the equation ${\cal P}_{\cal R} (k_*) = A_s (k_*)\vert _{\rm Planck}$ that yields compliance with all CMB observables.) As expected, by incorporating RG running of the SM parameters, we identify values of $\xi$ that are exponentially lower than the original estimate of $\xi \sim {\cal O} (10^4)$ from tree-level Higgs inflation. 

Although many points $\{ \lambda_{\rm ll}, \mu_{\rm m} , \delta, \xi \}$ exist that are consistent with both EW-scale measurements and CMB observations, however, for \emph{none} of these points would the model predictions remain insensitive to (unknown) UV physics. In Fig.~\ref{fig:proxiesdist} we show the cumulative distributions for the three unitarity proxies discussed in Sec.~\ref{sec:Unitarity} and defined in Eq.~(\ref{UVproxies}). In each case, we evaluate the value of the field-dependent quantities $F$, $h$, and $\chi$ at the time $t_*$ when CMB-scale perturbations exit the Hubble radius, for the same sets of parameters shown in Fig.~\ref{fig:xistarcompliance}. We then compare those values with the upper bound on the required scale of new physics, $\Lambda = E_{\rm max}^{\rm EW}$. A likely (but certainly not sufficient) requirement for model predictions to remain insensitive to the (unknown) new physics is if at least one of the ratios $\{F_*/\Lambda, h_* / \Lambda ,\chi_*/ \Lambda \}$ remains less than one. 

As shown in Fig.~\ref{fig:proxiesdist}, $\chi_* / \Lambda > 100$ for all cases, $h_* / \Lambda > 10$ for all cases, and $F_* / \Lambda > 3$ for all cases. In short, each unitarity proxy remains larger than unity across the entire region $\{ \lambda_{\rm ll}, \mu_{\rm m}, \delta, \xi \}$ that is (otherwise) consistent with both EW-scale measurements and CMB observations. The fact that each unitarity proxy exceeds one for field values relevant during inflation strongly suggests that any (unknown) UV physics that would restore perturbative unitarity would affect the model's predictions for CMB observables unless additional assumptions on its properties are invoked.

To determine whether this behavior is universal or merely a result of the SM input, we extended our parameter-space scan in the $\lambda_{\rm ll}$-$\mu_{\rm m}$ plane to include regions allowing for $\xi \sim \mathcal{O}(1)$. Our results are shown in Fig.~\ref{fig:proxiesdist2}. We find that even in this regime, in which $\Lambda \sim E_{\rm max}^{\rm EW} \sim M_{\rm Pl}$, demanding compliance with the CMB observables inevitably leads to unitarity proxies larger than one. In addition, these results reproduce the nontrivial dependence of model predictions in the vicinity of $\mu_{\rm m}\sim M_{\rm Pl}/\sqrt{\xi}$. Fig.~\ref{fig:proxiesdist2} demonstrates that there exist regions of parameter space that are consistent with CMB observations and for which (at least the most lenient) unitarity proxy $F$ yields $F_* / \Lambda < 1$; but all of these points lie exponentially far from the SM $3\sigma$ region. 

We have thus identified robust regions of parameter space for critical Higgs inflation in the metric-gravity formulation that are compliant with \emph{at most} two of the three criteria: EW-scale measurements, CMB observations, and insensitivity of model predictions to unknown UV physics.

\subsection{Results in Palatini Gravity}
\label{sub:Palatini_results}

We repeat the analysis of Sec.~\ref{sub:metric_results} for the case of critical Higgs inflation in the Palatini formulation of gravity, as defined in Sec.~\ref{sec:NearCriticalHiggs}.

\begin{figure}
    \centering
    \includegraphics[width=\linewidth]{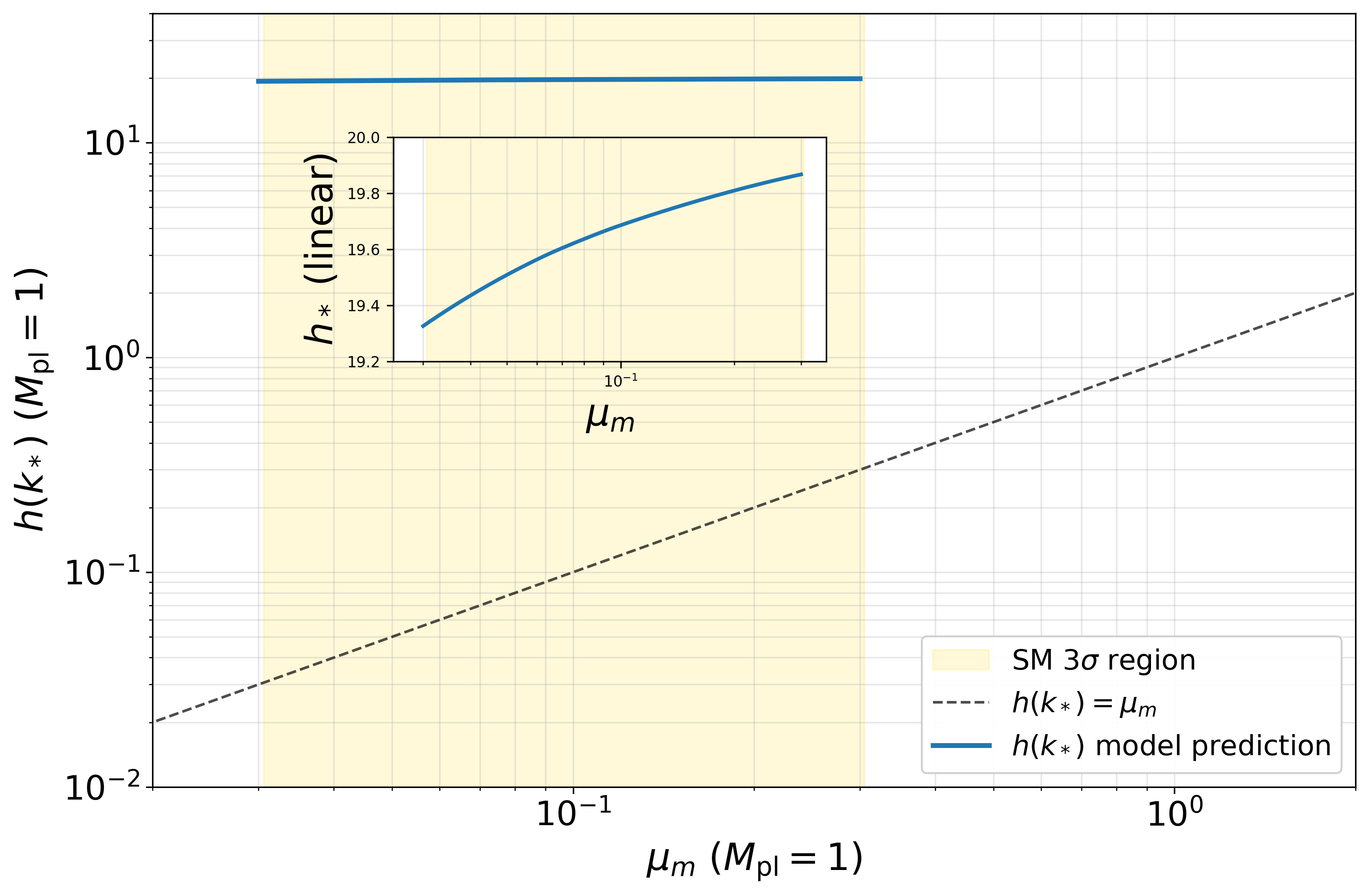}
    \caption{\justifying The value of the field $h_*$ at the time the CMB pivot scale exits the Hubble radius for (near-)critical Higgs inflation in the Palatini formulation, as a function of the inflection-point scale $\mu_{\rm m}$, for fixed $\lambda_{\rm ll} = 2\times 10^{-5}$ and $\delta = 0.01$, across the SM $3\sigma$ range in $\mu_{\rm m}$. The dashed line shows the value of the non-canonical field $h_*$ at the pivot time \textit{if the loop correction to $V(h)$, $\log^2 [z (h_*)]$, were vanishing}, which would yield $\lambda_{\rm eff}=\lambda_{\rm nl}$. By contrast, the blue curve shows $h_*$ across the full range of $\mu_{\rm m}$ values within the SM 3$\sigma$ region for the full two-loop RG-improved potential. The two-loop contribution to $\lambda_{\rm eff}$ ($\lambda_{\rm ll} \,\log^2 [z (h_*)]$) typically exceeds the tree-level term ($\lambda_{\rm nl}$) by 1-2 orders of magnitude. Each value of $\mu_{\rm m}$ along the blue curve corresponds to a different value of $\xi_*$, which is fixed by requiring that the predicted values of CMB observables $A_s, n_s, \alpha_s,\text{ and }r$ simultaneously lie within their  Planck $2\sigma$ bounds. The nonminimal coupling $\xi_*$ grows from $3.83\times10^{6}$ at $\mu_{\rm m} = 0.03 M_{\rm Pl}$ to $1.10\times 10^{7}$ at $\mu_{\rm m} = 0.3 M_{\rm Pl}$. 
  }
    \label{fig:h*vsmum}
\end{figure}

\begin{figure*}
    \centering
\includegraphics[width=0.8\linewidth]{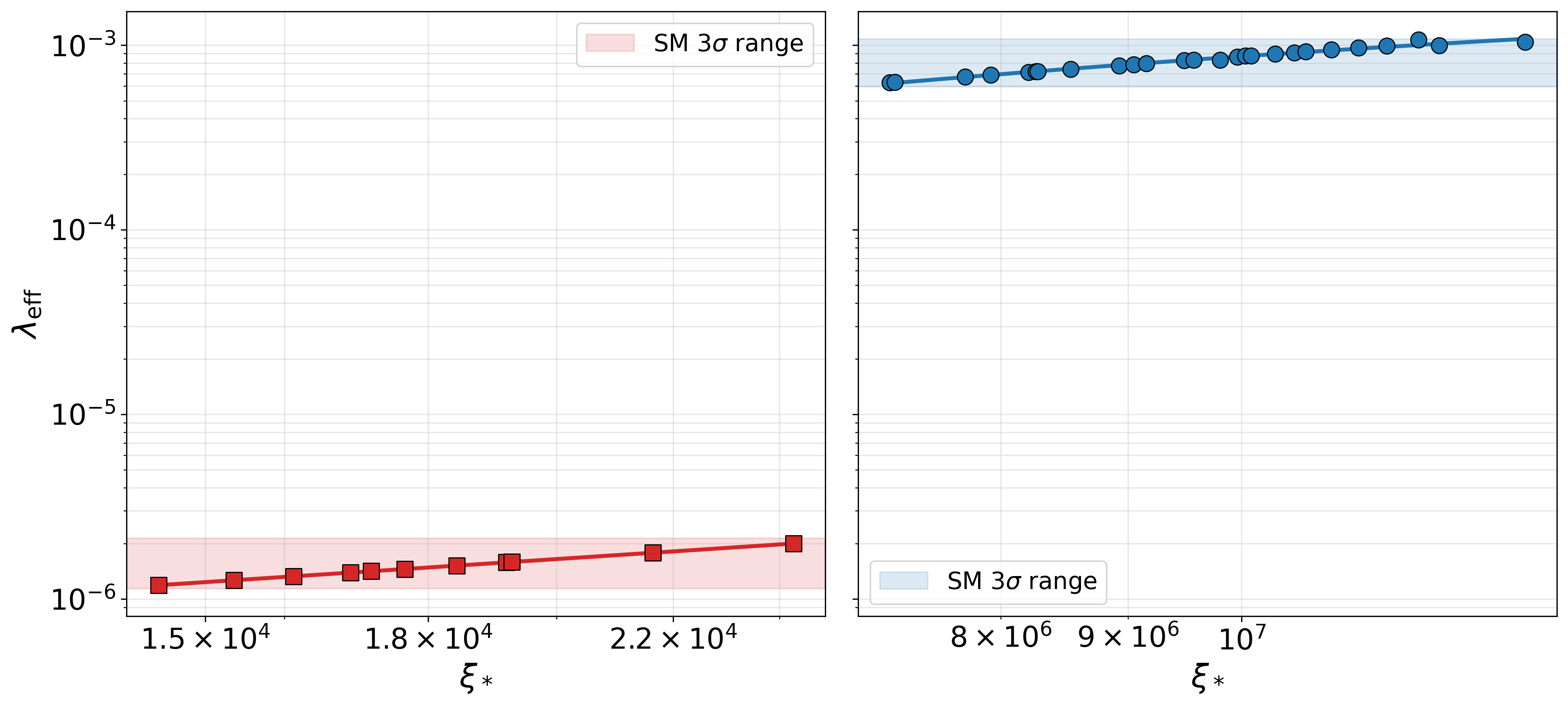}
    \caption{\justifying Nonminimal coupling $\xi_*$ versus effective quartic coupling $\lambda_{\rm eff}$ at the CMB pivot scale for Palatini-critical Higgs inflation with $\delta= 0.01$. The left-hand plot shows the tree-level result (red squares) with $\lambda_{\rm tree} \equiv \lambda_{\rm nl}$, while the right-hand plot shows the two-loop corrected result (blue circles), $\lambda_{\text{eff}} \equiv \lambda_{\rm nl} + \lambda_{\rm ll} \,\ln^2 [z (h_*)]$. Every plotted point is within the SM $3\sigma$ region while also matching all CMB observables to within $2\sigma$. The vertical offset between the tree-level (red, left) and two-loop (blue, right) curves makes clear the dominance of the loop terms over the tree-level contribution to the effective quartic at the CMB pivot scale, given that in the Palatini case $h_* \gg \mu_{\rm m}$.
  } 
    \label{fig:treeloopPalatini}
\end{figure*}

\begin{figure}
    \centering
\includegraphics[width=\linewidth]{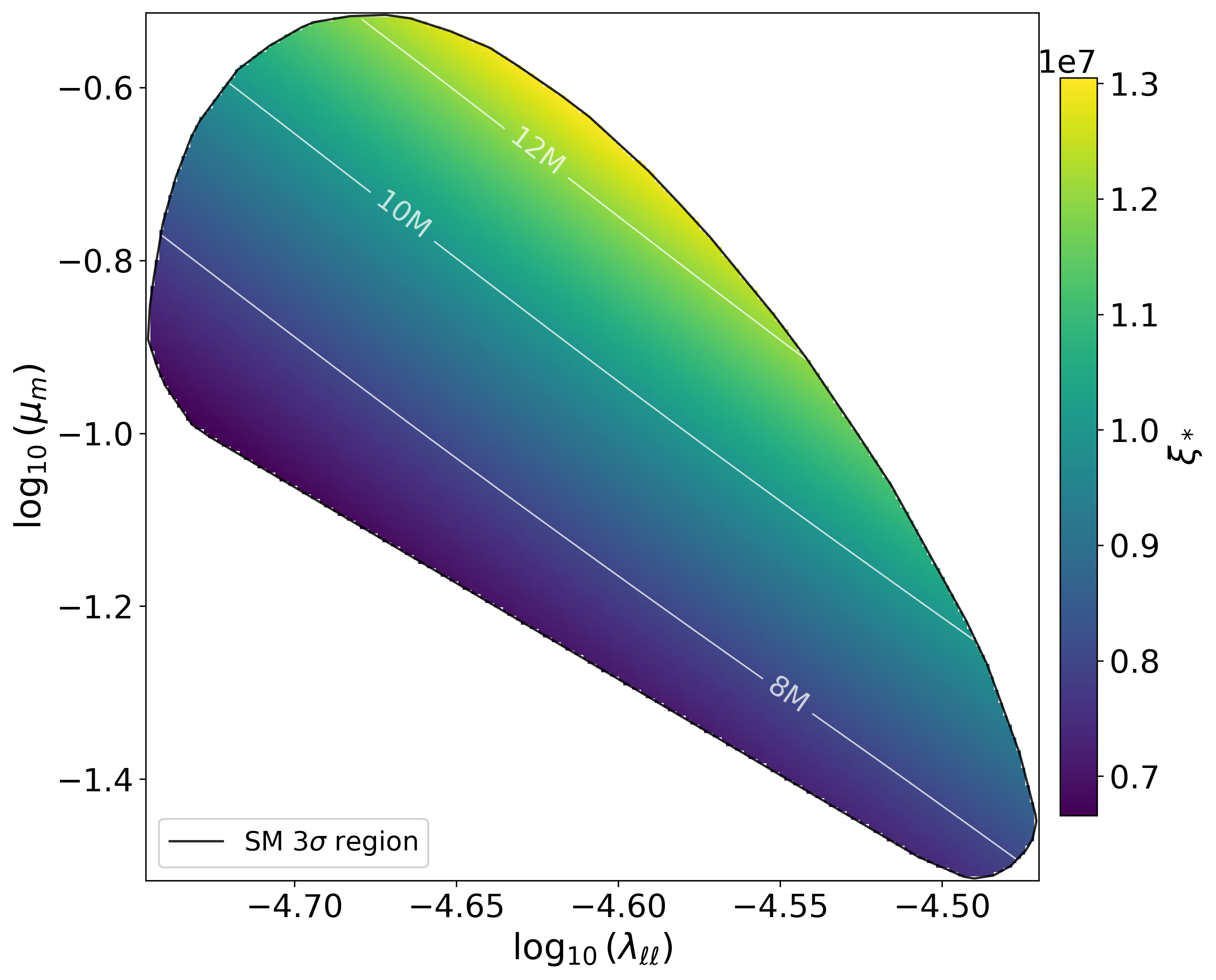}
\caption{\justifying The nonminimal coupling $\xi_*$ for Palatini-critical Higgs inflation as a function of $\mu_{\rm m}$ and $\lambda_{\rm ll}$, with $\delta = 0.01$. All shaded points lie within the SM $3\sigma$ region as well as within $2\sigma$ of the Planck CMB measurements. Values of $\xi_*$ are obtained with the root-finding algorithm  described in Appendix~\ref{sec:app_rootfinding}. Note that in the near-critical Palatini case, we require $\xi_* \sim {\cal O} (10^7)$ to maintain agreement with both EW-scale measurements and CMB observations.
}
\label{fig:palatini_topo}
\end{figure}

\begin{figure}
    \centering
\includegraphics[width=\linewidth]{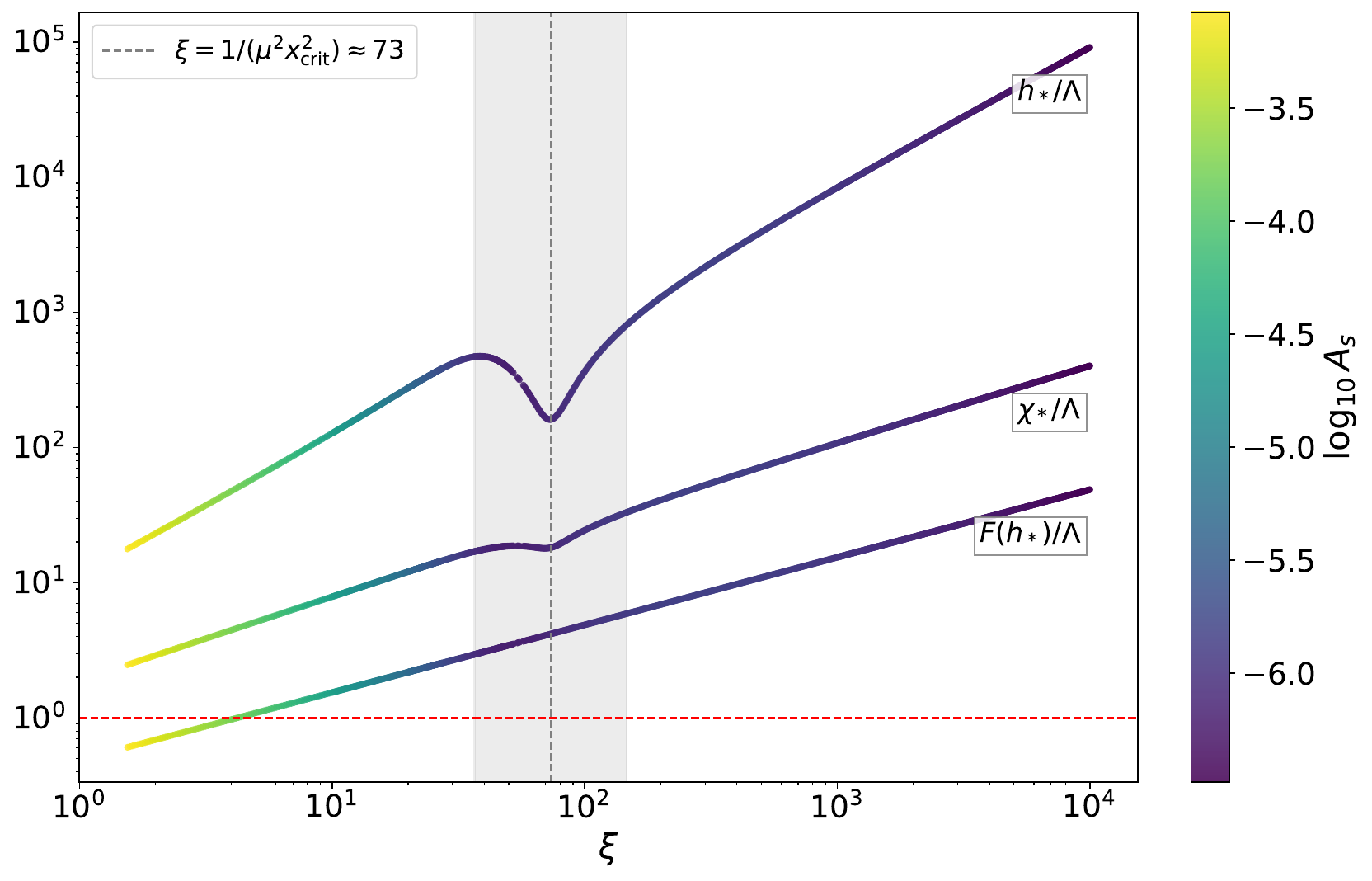}
    \caption{\justifying Dependence of the three UV-sensitivity proxies on the nonminimal coupling $\xi$ for a representative set of near-critical parameters in Palatini gravity. From top to bottom, the curves show $h_*/\Lambda$, $\chi_*/\Lambda$, and $F_*/\Lambda$, where $\Lambda \equiv E_{\max}^{\rm EW} = 4.94 M_{\rm Pl}/\sqrt{\xi}$, and the color bar to the right indicates the corresponding value of $\log_{10} A_s (k_*)$. The horizontal red dashed line marks the perturbative-unitarity cutoff. When the proxies approach unity, the predicted amplitude of primordial curvature perturbations at CMB scales is $A_s \sim 10^{-3}$, six orders of magnitude larger than the observed value. Matching the measured CMB amplitude therefore requires substantially larger values of $\xi$, for which all three proxies lie far above the unitarity scale. Each curve exhibits a kink around the value of $\xi(\mu_{\rm m})$ which marks the transition between the regime where the inflection point of the near-critical potential is near the end of inflation and where it is high up on the plateau, relevant for CMB scales. 
    As also shown in Fig.~\ref{fig:6panel}, this transition occurs at $\mu_{\rm m}^2 z_{\rm crit}^2\sim M_{\rm Pl}^2/\xi_{\rm crit}$, where $z_{\rm crit}=e^{-1/4}$ which, for the fixed value of $\mu_{\rm m} = 0.15 \, M_{\rm Pl}$ shown in this plot, lies at $\xi_{\rm crit}\approx 73$.  }
    \label{fig:PalatiniUnitarityProxies}
\end{figure}

The differences between the Palatini and metric-gravity cases are reflected in the distinct mappings between the Higgs field $h$ and the canonically normalized field $\chi (h)$, as given by Eqs.~(\ref{chimetric}) and (\ref{chiPalatini}) and illustrated in Fig.~\ref{fig:HiggsChiMaps}. For a given value of $\xi$, $h (\chi)$ rises much sharper with $\chi$ in the Palatini case than in the metric case. In particular, a typical inflationary excursion $\Delta \chi \sim M_{\rm Pl}$ corresponds to $\Delta h \sim M_{\rm Pl} / \sqrt{\xi}$ in the metric-gravity case, and to $\Delta h \sim M_{\rm Pl} \, e^{\sqrt{\xi}}/ \sqrt{\xi}$ in the Palatini-gravity case.

This steep scaling of $h$ with $\chi$ in the Palatini-gravity case means that the effective potential $V_E (\chi)$ has an even flatter plateau at large $\chi$ than in the metric-gravity case. The smaller value of $V_E^\prime (\chi)$, in turn, leads to a smaller value of the slow-roll parameter $\epsilon$ while the field evolves along the large-field plateau. Given the scaling ${\cal P}_{\cal R} (k_*)= A_s (k_*) \sim V_E (\chi_*) / \epsilon (\chi_*)$ in Eq.~(\ref{PRSRV}), the suppressed value of $\epsilon (\chi_*)$ in the Palatini case (compared to the metric-gravity case) pushes the amplitude $A_s (k_*)$ to larger values, relative to comparable points in parameter space for the metric-gravity case. This behavior is well-known from studies of tree-level Palatini-Higgs inflation, for which one requires exponentially larger values of $\xi$ to bring $A_s (k_*)$ into compliance with CMB measurements: 
\begin{equation}
\xi_{\rm tree, Palatini} \sim {\cal O} (10^5) \times \left(\frac{\lambda_{\rm eff} }{ 10^{-5} } \right)
\end{equation}
for the Palatini case, compared to 
\begin{equation}
\xi_{\rm tree, metric} \sim {\cal O} (10^2) \times \left( \frac{ \lambda_{\rm eff} }{ 10^{-5} } \right)^{1/2}
\end{equation}
for the metric case \cite{Rubio:2018ogq,Steingasser:2025txd}.

This difference becomes magnified in the case of critical Higgs inflation, for which $\lambda_{\rm eff}$ receives contributions from the two-loop corrections. Near the critical point, the effective potential in the Einstein frame behaves as in Eq.~(\ref{eq:VE}), with
\begin{equation}
    \lambda_{\rm eff} = \lambda_{\rm nl} + \lambda_{\rm ll} \,\ln^2 [ z (h) ] ,
    \label{lambdaeffcrit}
\end{equation}
with $\lambda_{\rm nl} \simeq \lambda_{\rm nl, crit} = \lambda_{\rm ll} / 16$ from Eq.~(\ref{lambdanlcrit}), and using the definition of $z(h) = F(h) / \mu_{\rm m}$ from Eqs.~(\ref{eq:zdef}) and (\ref{Fdef}). 

As shown in Fig.~\ref{fig:h*vsmum}, when we consider near-critical Higgs inflation in the Palatini formulation, the large values of the (non-canonical) Higgs field $h_*$ at the time the CMB pivot scale exits the Hubble radius generically make the logarithmic corrections to $\lambda_{\rm eff}$ in Eq.~(\ref{lambdaeffcrit}) dominate over the tree-level contribution.\footnote{Note that this does \textit{not} imply a breakdown of the loop expansion more generally, but only the subdominance of $\lambda_{\rm nl}$. Near-criticality requires $\lambda_{\rm nl}\simeq \lambda_{\rm ll}/16$, such that $\ln [z (h_*)]=\mathcal{O}(1)$ is sufficient for $\lambda_{\rm eff}$ to be dominated by its two-loop contribution. Near-criticality does \textit{not} affect the power-counting of higher orders in the loop expansion, which remain suppressed by additional powers of the SM couplings and the usual loop factor of $(4 \pi)^{-2}$.} This, in turn, means that for given values of $\{ \lambda_{\rm ll}, \mu_{\rm m}, \delta \}$, the effective coupling $\lambda_{\rm eff}$ becomes much {\it larger} than in the simple, tree-level case, as shown in Fig.~\ref{fig:treeloopPalatini}. 

The increased value of $\lambda_{\rm eff}$ requires a larger value of $\xi$ in order for the amplitude of the primordial spectrum $A_s (k_*)$ to match the measured CMB value, compared to the tree-level Palatini case; see Fig.~\ref{fig:palatini_topo}. The larger values of $\xi \gtrsim {\cal O} (10^7)$ required to match CMB measurements reduce the perturbative unitarity scale $\Lambda = E_{\rm max}^{\rm EW} = 4.94 \, M_{\rm Pl} / \sqrt{\xi}$, pushing each of the UV-sensitivity proxies ($h_* / \Lambda, \chi_* / \Lambda$, and $F(h_*) / \Lambda$) to be much greater than 1. See Fig.~\ref{fig:PalatiniUnitarityProxies}. 

Much as in the case for critical Higgs inflation in metric gravity, we therefore find that critical-Palatini Higgs inflation can satisfy at most two of our three criteria: consistency with EW-scale measurements, consistency with CMB observations, and limited sensitivity to unknown UV physics.  

\section{Discussion}
\label{sec:Discussion}
We have revisited critical Higgs inflation based on three criteria: consistency with EW-scale measurements, agreement with CMB observables, and insensitivity of those predictions to the unknown physics required to restore perturbative unitarity. Within the near-critical parameter space considered here, we find no region in which all three criteria are satisfied simultaneously, in either the metric or Palatini formulation of gravity.

The first two criteria can be simultaneously satisfied. EW-scale measurements constrain the parameters $\mu_m$ and $\lambda_{\rm ll}$ through their RG flow, while the value of $\lambda_{\rm nl}$ is especially sensitive to small changes near criticality. We therefore allowed a threshold correction to $\lambda_{\rm nl}$, parameterized by $\delta$, and considered an extended region of parameter space consistent with the measured SM couplings. For each viable choice of the potential parameters, the nonminimal coupling $\xi$ can be chosen so that the predicted CMB observables agree with measurements. Thus, the difficulty we identify does not arise from an inability of the near-critical potential to support inflation or reproduce the observed scalar power spectrum.

In metric gravity, matching the CMB observables within the SM-compatible region requires $50\lesssim\xi\lesssim400$ for the representative near-critical choice $\delta=0.01$. These values are substantially smaller than those usually required in tree-level Higgs inflation ($\xi \sim {\cal O}(10^4)$). Nevertheless, for every corresponding parameter point in our scan, all three proxies in Eq.~\eqref{UVproxies} exceed the EW-vacuum unitarity scale $E^{\rm EW}_{\rm max} = 2.05 \, M_{\rm Pl} / \xi$: $F_*/E^{\rm EW}_{\rm max}>3$, $h_*/E^{\rm EW}_{\rm max}>10$, and $\chi_*/E^{\rm EW}_{\rm max}>100$. Extending the scan to values of $\xi$ of order unity reveals CMB-compatible points with $F_*/\Lambda<1$, but these lie well outside the region inferred from EW-scale measurements. Reducing $\xi$ therefore does not resolve the tension among the three criteria.

The same tension persists in Palatini gravity, for a different reason. The relation between $h$ and the canonically normalized field $\chi$ places the CMB pivot scale at much larger values of $h$ than in the metric case. Near criticality, the constant contribution $\lambda_{\rm nl}$ to the effective quartic coupling is suppressed, so the logarithmic term in Eq.~\eqref{lambdaeffcrit} becomes important over the field range relevant for the CMB. Matching the observed scalar amplitude then requires $\xi \sim {\cal O}(10^6 - 10^7)$ across the SM-compatible region examined here. At those values, the three proxies again lie above the EW-vacuum unitarity scale.

The cutoff used in Eq.~\eqref{lambdaeffcrit} follows from scattering around the EW vacuum, whereas the background-dependent scale obtained by expanding around the inflationary condensate can be much higher \cite{Steingasser:2025txd}. Our comparison tests whether one effective action can connect the measured Higgs sector at EW scales to inflation without predictions becoming sensitive to additional, unspecified terms. The field ratios are proxies for that sensitivity; their values alone do not determine the effects of a particular UV completion. Indeed, the small effective quartic coupling near criticality makes the potential especially susceptible to higher-dimensional contributions, as illustrated by Eq.~\eqref{eq:sensest}, even when a field ratio is below unity.

Critical Higgs inflation may therefore remain viable within a specified UV completion. Establishing that possibility would require identifying the new physics that restores unitarity, determining its effects on the RG flow and the effective action, and calculating the resulting inflationary observables. Such a calculation would also be necessary to assess proposed small-scale features of the primordial spectrum, including those associated with enhanced gravitational waves or primordial black holes. In the absence of that additional information, agreement with CMB measurements does not yield predictions that can be attributed to the measured SM Higgs sector alone.

\section*{Acknowledgments}
It is a pleasure to thank Evan McDonough, Jos\'{e}-Ramon Espinosa, Benjamin V.~Lehmann, Wenzer Qin, Shyam Balaji, and Javier Rubio for helpful discussions. Portions of this work were conducted in MIT's Center for Theoretical Physics---A Leinweber Institute and partially supported by the U.S. Department of Energy under Contract No. DE-SC0012567. SRG is supported by an NSF Mathematical and Physical Sciences Ascend Fellowship  under Award No.~2317018. TS acknowledges financial support from the Spanish Research Agency (Agencia Estatal de Investigaci\'on) through the grant IFT Centro de Excelencia Severo
Ochoa No CEX2020-001007-S and PID2022-137127NBI00 funded by MCIN/AEI/10.13039/501100011033/FEDER, UE. TS also acknowledges financial support by the Spanish Research Agency (Agencia Estatal de Investigaci\'on) through grant PID2025-172338NB-I00, funded by MICIU/AEI/10.13039/501100011033 and by ERDF, EU. This project has received funding/support from the European Union's Horizon 2020 research and innovation programme under the Marie Sklodowska-Curie Staff Exchange grant agreement No. 101086085-ASYMMETRY. 
The opinions expressed in this publication are those of the authors and do not necessarily reflect the views of these Foundations. 

\appendix

\section{Criticality condition}
\label{App:Criticality}
\begin{figure}[h!]
    \centering
    \includegraphics[width=\linewidth]{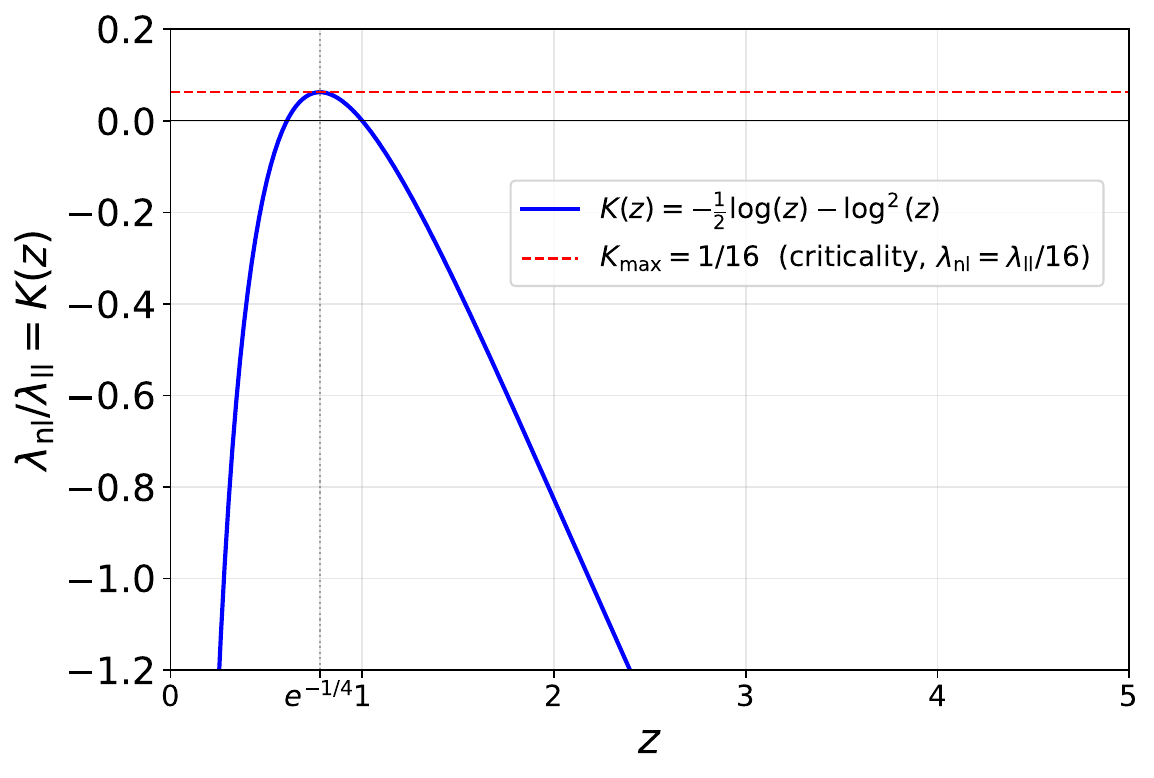}
    \caption{\justifying The function $K(z)$ as introduced in Eq.~(\ref{lambdaratio}). The maximum value lies at $z=e^{-1/4}$ as indicated by the intersection of the vertical dotted gray line and the horizontal red dashed line.}
    \label{fig:F(z)}
\end{figure}
In Sec.~\ref{sec:NearCriticalHiggs}, we found that the Higgs potential (including effects of the nonminimal coupling) can be brought to the form
\begin{align}
    V_{E}(h, \mu_{\rm m})=\frac{1}{4} &\left[ \lambda_{\rm nl}+ \lambda_{\rm ll} \cdot  \ln^2 \left(z(h)\right)  \right] \cdot  z^4(h) \cdot \mu_{\rm m}^4,
\end{align}
where $z (h)$ is defined in Eq.~(\ref{eq:zdef}).

Moreover, we identified the criticality condition by considering the transition from the quantum phase corresponding to a potential that ``flips over'' to one in which it is monotonically rising. This distinction can be quantified through the number of extrema in the potential: ignoring the mass term, the first of these phases is characterized by a total of three local extrema, while the latter only gives rise to a single minimum at $h=0$. In other words, these phases are characterized by the number of solutions to the equation
\begin{align}
    0=\frac{d}{d z} V_E (z)= \left[ \frac{\lambda_{\rm nl}}{\lambda_{\rm ll}} + \frac{1}{2} \log (z) + \log^2 (z) \right] \lambda_{\rm ll} \mu_{\rm m}^4 z^3.
\end{align}
Note that this definition is unambiguous due to its invariance under field redefinitions, which would merely contribute a multiplicative factor. Dropping the constant factors and focusing on the non-trivial case $z\neq 0$, this equation can be rewritten as
\begin{align}
    \frac{\lambda_{\rm nl}}{\lambda_{\rm ll}} =-  \frac{1}{2} \log (z) - \log^2 (z) \equiv K (z).
    \label{lambdaratio}
\end{align}
This relation can easily be understood by observing that the function $K(z)$ is bounded from above for real values of $z$, $K(z)\leq 1/16$. Thus, for $\lambda_{\rm nl}/\lambda_{\rm ll}>1/16$, there exists no real, non-vanishing solution for Eq.~(\ref{lambdaratio}). If, however, $\lambda_{\rm nl}/\lambda_{\rm ll} < 1/16$, two such solutions exist. We therefore recover the criticality condition $\lambda_{\rm nl,crit}=\lambda_{\rm ll}/16$. See Fig.~\ref{fig:F(z)}.

Next, we can determine the position of the inflection point, which we denote as $z_{\rm crit}\equiv z(h_{\rm crit})$. For $\lambda_{\rm nl}=\lambda_{\ell \ell}/16$, Eq.~\eqref{lambdaratio} reduces to 
\begin{gather}
    \frac{1}{16} =-  \frac{1}{2} \log (z_{\rm crit}) - \log^2 (z_{\rm crit}) \\ \Rightarrow \log (z_{\rm crit})=-\frac{1}{4}.
\end{gather}
Thus, the inflection point forms at
\begin{align}
    z_{\rm crit}=e^{-1/4} \quad \Leftrightarrow \quad h_{\rm crit}= \frac{\mu_{\rm m} }{\sqrt{e^{1/2}-\xi \mu_{\rm m}^2/M_{\rm Pl}^2}}.
\end{align}

\section{Expressing potential parameters through RG quantities}
\label{appendixBetaFunctions}

For parameters close to a critical point, it is convenient to express the parameters in the potential of Eq.~\eqref{Veff1} in terms of the running coupling $\lambda(\mu)$. This can be achieved by making use of the Callan-Symanzik equations for the effective potential
\begin{align}
    0=&\mu \frac{d}{d \mu}  V_{\rm eff}(\mu, \lambda(\mu), h(\mu)) \\ 
    =&\left[ \mu \frac{\partial}{\partial \mu}  + \sum_i \beta_{\lambda_i} \frac{\partial}{\partial \lambda_i}  - \gamma h \frac{\partial}{\partial h} \right] V_{\rm eff}(\mu, \lambda, h) \nonumber
\end{align}
Using the concrete expression for the potential in Eq.~\eqref{Veff1}, it is straightforward to evaluate this equation at two-loop accuracy,
\begin{align}
    0=&\beta_\lambda - \lambda_{\rm l} - \gamma \lambda_{\rm l} - 4 \gamma \lambda_{\rm nl} \\
    &+ \left(\beta_{\lambda_{\rm l}} -  4 \gamma \lambda_{\rm l} -2 \lambda_{\rm ll} -2 \gamma \lambda_{\rm ll}\right)  \log\left(\frac{h}{\mu}\right),
\end{align}
where each line needs to hold separately. Near $\mu_{\rm m}$, for which $\lambda_{\rm l}(\mu_{\rm m})=0$ and $\lambda_{\rm nl}(\mu_{\rm m})\sim \mathcal{O}$(two-loop), the first equation allows us to relate $\mu_{\rm m}$ to the beta function $\beta_\lambda$,
\begin{align}\label{eq:CSl}
    \beta_\lambda (\mu_{\rm m}) = \lambda_{\rm l} (\mu_{\rm m}) (1 + \gamma (\mu_{\rm m}) ) +  4 \gamma  (\mu_{\rm m})  \lambda_{\rm nl}  (\mu_{\rm m}) \simeq 0
\end{align}
Thus, we can indeed identify $\mu_{\rm m}$ with the RG scale at which $\lambda$ reaches its minimum. We can extract an additional relation by considering the terms $\propto \log(h/\mu)$, 
\begin{align}
   0= \beta_{\lambda_{\rm l}} - 4 \gamma \lambda_{\rm l} -2 \lambda_{\rm ll} -2 \gamma \lambda_{\rm ll}
\end{align}
We can again simplify this at $\mu_{\rm m}$, yielding a straightforward expression for $\lambda_{\rm ll}$:
\begin{align}
   \lambda_{\rm ll} (\mu_{\rm m})=\frac{1}{2}\beta_{\lambda_{\rm l}} (\mu_{\rm m}) \simeq \frac{1}{2} \left( \mu \frac{\partial \beta_{\lambda}}{\partial \mu} \right)\bigg|_{\mu = \mu_{\rm m}}\hspace{-3mm} \equiv \frac{1}{2} \beta_\lambda^\prime (\mu_{\rm m}) .
\end{align}
In the last step, we have used Eq.~\eqref{eq:CSl} at two-loop level.

\section{Conformal transformation in metric and Palatini formulations}
\label{App:ConformalTransformation}

In Eq.~\eqref{eq:GIJ} we saw that conformally transforming the Jordan action leads to an Einstein-Hilbert action in which the kinetic sector differs between the metric and Palatini formulations of gravity. In particular, depending on whether one considers the connection to vary independently of the metric, one finds the induced field-space curvature in the Einstein frame to take the form
\begin{gather}
     {\cal G}_{IJ}= \frac{M_{\rm Pl}^2}{M_{\rm Pl}^2+ \xi h^2}\left[ \delta_{IJ} + 6 \sigma \xi^2 \frac{\delta_{IK} \delta_{JL} \phi^K \phi^L}{M_{\rm Pl}^2+ \xi h^2} \right],
\end{gather}
where $\sigma=1$ in metric gravity and $\sigma=0$ in Palatini gravity, and $h^2 = 2 \vert {\bf H} \vert^2$.

To compute the transformed action in each scheme, we begin with the conformal transformation $\Omega^2(x)$ satisfying 
\begin{equation}
g_{\mu\nu}(x) = \Omega^2(x)\, \tilde{g}_{\mu\nu}, 
\label{conftrans1}
\end{equation}
from which it follows that
\begin{equation}
\begin{split}
    g^{\mu \nu} &= \frac{1}{\Omega^2(x)} \tilde{g}^{\mu\nu} \\
    \sqrt{-g} &= \Omega^4(x) \sqrt{-\tilde{g}}.
\end{split}
\label{ginverseconf}
\end{equation}
To bring the gravitational portion of the action into the usual Einstein-Hilbert form, we select $\Omega(x)$ to be
\begin{gather}
        \Omega^2(x) = \frac{2f[h(x)]}{ M_{\rm Pl}^2},
        \label{Omegadef2}
\end{gather}
where the nonminimal coupling function $f(h)$ is defined according to Eq.~\eqref{eq:NMcoup} as
\begin{equation}
      f[h(x)] = \frac{1}{2}(M_{\rm Pl}^2 +\xi h^2(x)).
\end{equation}
Recall from Eqs.~(\ref{Hdoublet}) and (\ref{Hchi}) that $h^2 = \delta_{IJ} \, \phi^I \phi^J$.

\subsection{Metric Gravity}

The action in the Jordan frame for the metric-gravity case is \cite{Kaiser:2010ps,Schutz:2013fua}:
\begin{gather}
    S = \int d^4 x \sqrt{-\tilde{g}}\left[ f(\phi^I) \tilde{R} - \frac{1}{2} \tilde{g}^{\mu\nu} \delta_{IJ} \,\partial_{\mu} \phi^I \partial_{\nu} \phi^J  - \tilde{V}(\phi^I) \right].
\label{SJ}
\end{gather}
For metric gravity, the connection $\Gamma^\alpha_{\beta\gamma}$ is taken to be a function of $g_{\mu\nu}$ and $\partial_\alpha g_{\mu\nu}$. Therefore, under the conformal transformation of Eq.~(\ref{conftrans1}), the connection transforms as (see, e.g., Ref.~\cite{Kaiser:2010ps})
\begin{equation}
    \Gamma^\alpha_{\beta\gamma} = \tilde{\Gamma}^\alpha_{\beta\gamma} + \frac{1}{ \Omega} \Big[ \delta^\alpha_\beta \nabla_\gamma \Omega + \delta^\alpha_\gamma \nabla_\beta \Omega - g_{\beta \gamma} \nabla^\alpha \Omega \Big] ,
\label{Gammatransform}
\end{equation}
and in $D$ spacetime dimensions, the Ricci scalar transforms as \citep{Kaiser:2010ps}:
\begin{equation}
\begin{split}
R (g) &= \frac{1}{ \Omega^2} \left[ \tilde{R} (\tilde{g})- \frac{ 2 (D - 1)}{\Omega} \widetilde{\Box} \Omega \right. \\
&\left. \quad\quad\qquad- \frac{ (D - 1)(D - 4)}{\Omega^2} \tilde{g}^{\mu\nu} \tilde{\nabla}_\mu \Omega \tilde{\nabla}_\nu \Omega \right] .
    \label{eq:ricciM}
\end{split}
\end{equation}
Upon making use of Eqs.~(\ref{ginverseconf}), (\ref{Omegadef2}), and (\ref{eq:ricciM}), and performing an integration by parts, the action of Eq.~(\ref{SJ}) may be written \cite{Kaiser:2010ps}
\begin{equation}
\begin{split}
S 
= \int d^4 x \sqrt{-g} &\left[ \frac{ M_{\rm pl}^2}{2} R (g) - \frac{3 M_{\rm pl}^2}{\Omega^2} g^{\mu\nu} \nabla_\mu \Omega \nabla_\nu \Omega \right.\\
&\>\>\left. -  \frac{1}{2 \Omega^2} g^{\mu\nu} \delta_{IJ} \,\partial_{\mu}\phi^I \,\partial_{\nu}\phi^J - \frac{\tilde{V}(\phi^I)}{\Omega^4}  \right]
\label{eq:EH}
\end{split}
\end{equation}
for $D = 4$. Using Eq.~(\ref{Omegadef2}) for $f (\phi^I)$ yields
\begin{equation}
    \int d^4 x \sqrt{-g} \left[ \frac{M_P^2}{2} R(g) -
    \frac{1}{2} g^{\mu\nu} {\cal G}_{IJ}\partial_{\mu}\phi^I \partial_{\nu}\phi^J -
    V(\phi)\right] ,
\end{equation}
where $V (\phi^I) \equiv \tilde{V} (\phi^I) / \Omega^4$ and ${\cal G}_{IJ} (\phi^K)$ matches Eq.~\eqref{eq:GIJ}, with $\sigma=1$.

\subsection{Palatini Gravity}

In the Palatini formulation of gravity, we treat the metric and connection as independent variables. We again begin with the action in the Jordan frame,
\begin{gather}
     S = \int d^4 x  \sqrt{-\tilde{g}} \left[ f(\phi^I) \tilde{R}_{\rm P} - \frac{1}{2}\tilde{g}^{\mu\nu} \delta_{IJ} \,\partial_\nu \phi^I \partial_\mu^J \phi  - \tilde{V}(\phi) \right].
    \label{SJP}
\end{gather}
The object $\tilde{R}_{\rm P}$ is the Ricci scalar in the Palatini formulation, which obeys $\tilde{R}_{\rm P} = \tilde{g}^{\mu\nu} \tilde{R}_{\mu\nu}(\Gamma)$, where $\tilde{R}_{\mu\nu}(\Gamma)$ is the Ricci tensor. Under the conformal transformation of Eq.~(\ref{conftrans1}), the Ricci tensor $\tilde{R}_{\mu\nu} (\Gamma)$ remains unchanged, while the Ricci scalar transforms as
\begin{equation}
   \tilde{R}_{\rm P} = \tilde{g}^{\mu\nu} \tilde{R}_{\mu\nu}(\Gamma) = \Omega^2 g^{\mu\nu} R_{\mu\nu}(\Gamma).
\end{equation}
Then the action of Eq.~(\ref{SJP}) transforms into
\begin{equation}
\begin{split}
    S &= \int d^4 x  \frac{\sqrt{-g}}{\Omega^4}\,\, \Big[ f(\phi) \, \Omega^2 g^{\mu\nu} R_{\mu\nu}(\Gamma)  \\
    & \qquad\qquad \qquad\qquad - \frac{\Omega^2}{2} \, g^{\mu\nu} \delta_{IJ} \,\partial_\mu \phi^I \partial_\nu \phi^J  - \tilde{V}(\phi) \Big]
    \\
    &=  \int d^4x \sqrt{-g} \left[\frac{M_{\rm Pl}^2}{2}R_{\rm P} 
- \frac{1}{2} g^{\mu\nu}\mathcal{G}_{IJ}\partial_\mu \phi^I \partial_\nu \phi^J  - V(\phi) \right] ,
\end{split}
\end{equation}
where ${\cal G}_{IJ} (\phi^K)$ matches Eq.~\eqref{eq:GIJ} with $\sigma=0$.

\section{An algorithmic root-finding method for identifying CMB-compliant parameter sets}\label{sec:app_rootfinding}
\begin{figure*}
    \centering
    \includegraphics[width=0.8\linewidth]{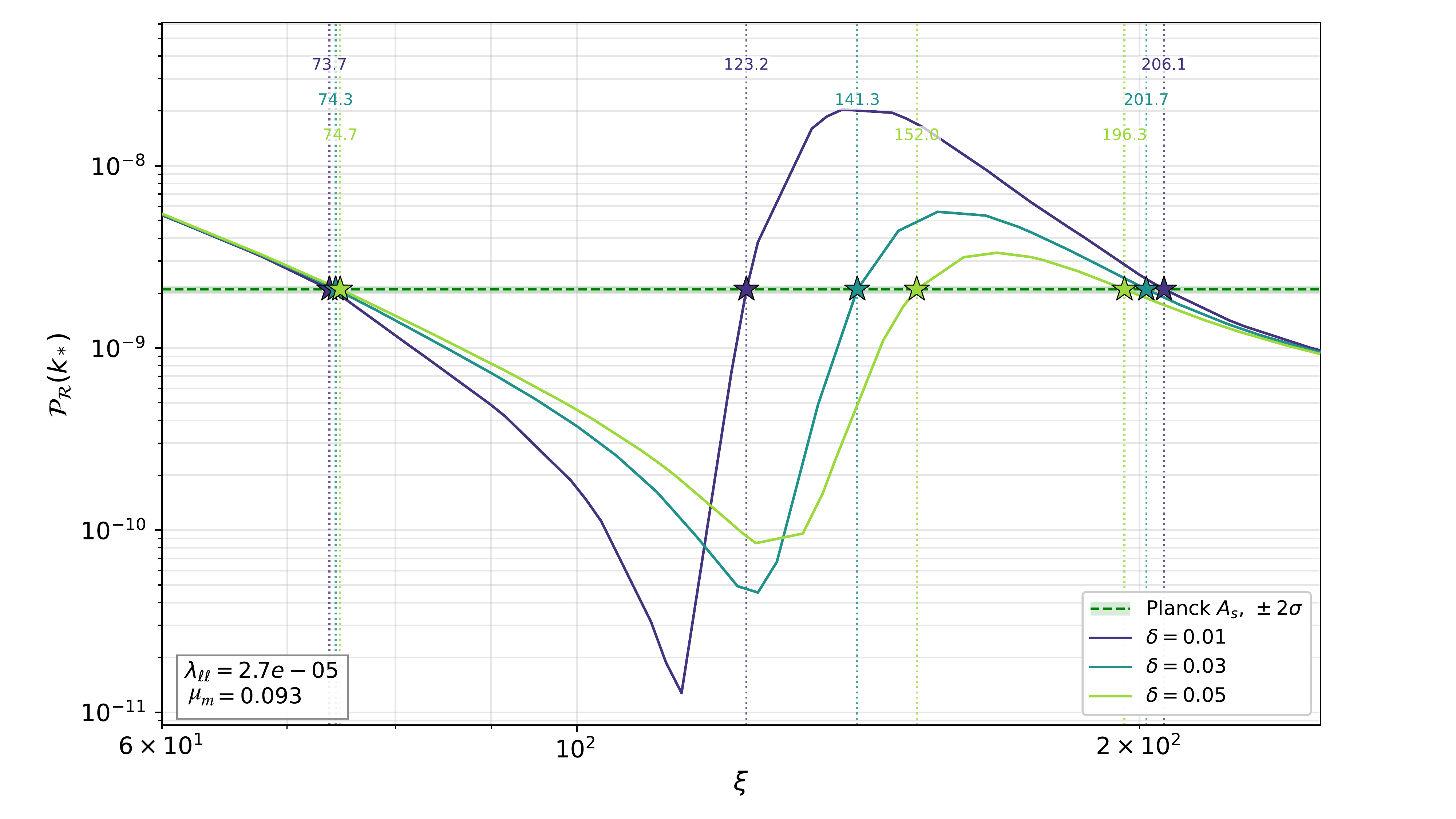}
    \caption{\justifying Dimensionless power spectrum of the scalar curvature perturbations at the CMB pivot scale, $\mathcal{P}_{\mathcal{R}}(k_*) = A_s (k_*)$, as a function of the nonminimal coupling $\xi$, for the benchmark parameters $\lambda_{\rm ll} = 2.7 \times 10^{-5}$ and $\mu_{\rm m} = 0.093 M_{\rm Pl}$, and for $\delta = 0.01$, $0.03$, and $0.05$. The horizontal green line and accompanying $2\sigma$ interval indicate the measured value of $A_s$. Stars mark intersections with the central observed value, and the corresponding vertical dotted lines and labels identify the roots $\xi_*$. For each value of $\delta$, the nonmonotonic dependence of $\mathcal{P}_{\mathcal{R}}(k_*)$ on $\xi$ produces three roots, illustrating why a global bracketing procedure is required to identify all CMB-compatible solutions.}
\label{fig:roots_for_three_deltas}
\end{figure*}
As discussed in Sec.~\ref{sec:NumericalResults}, a typical subset of parameters $\{\lambda_{\rm ll}, \mu_{\rm m}, \delta \}$ for $\lambda_{\rm ll}$ and $\mu_{\rm m}$ within the $3\sigma$ region set by EW precision data and with $0 \lesssim \delta \lesssim 0.5$ will match predicted values for most CMB observables fairly well for some value of $\xi \sim {\cal O} (10-10^3)$, chosen so as to match the measured amplitude of the power spectrum of primordial curvature perturbations. Very few parameter choices with $\{\lambda_{\rm ll}, \mu_{\rm m} \}$ within the SM $3\sigma$ region produce a tensor-to-scalar ratio above the current observational bounds, but we predictably find that choices of $\xi$ for which $\mu_{\rm m}\sim M_{\rm Pl} / \sqrt{\xi}$ lead to values of $r > 0.036$, as discussed in Sec.~\ref{sec:NumericalResults}.

Since the scaling of ${\cal P}_{\cal R}(k_*)$ with model parameters is significantly more complicated in the critical Higgs model than in tree-level Higgs inflation---especially once loop corrections introduce further correlations between SM and CMB parameters through the combination $F(h_* (\xi))/\mu_{\rm m}$---there is no simple analytic prescription for determining the value of $\xi$ for a given set of values of $\{\lambda_{\rm ll},\mu_{\rm m}, \delta\}$ that yields \textit{both} ${\cal P}_{\cal R}(k_*)=A_s=2.09\times 10^{-9}$ and $n_s, \alpha_s$, and $r$ within 2$\sigma$ of their Planck central values. Instead, $A_s(\xi)$ depends on $\xi$ in a nonlinear way through the potential as well as through the field values, determined by the background dynamics, which themselves depend on $\xi$. 

Here we give a reproducible numerical algorithm which, applied to a grid of points masked by the SM $3\sigma$ region, yields a full mapping of the $\{\lambda_{\rm ll}, \mu_{*}, \xi \}$ parameter space \emph{for any fixed value of $\delta$}. The algorithm can then be repeated for different values of $\delta$ as desired. The result of one such scan at $\delta=0.01$ is plotted in Fig.~\ref{fig:chickenpoxplot}. 
\begin{figure*}
    \centering
    \includegraphics[width=0.7\linewidth]{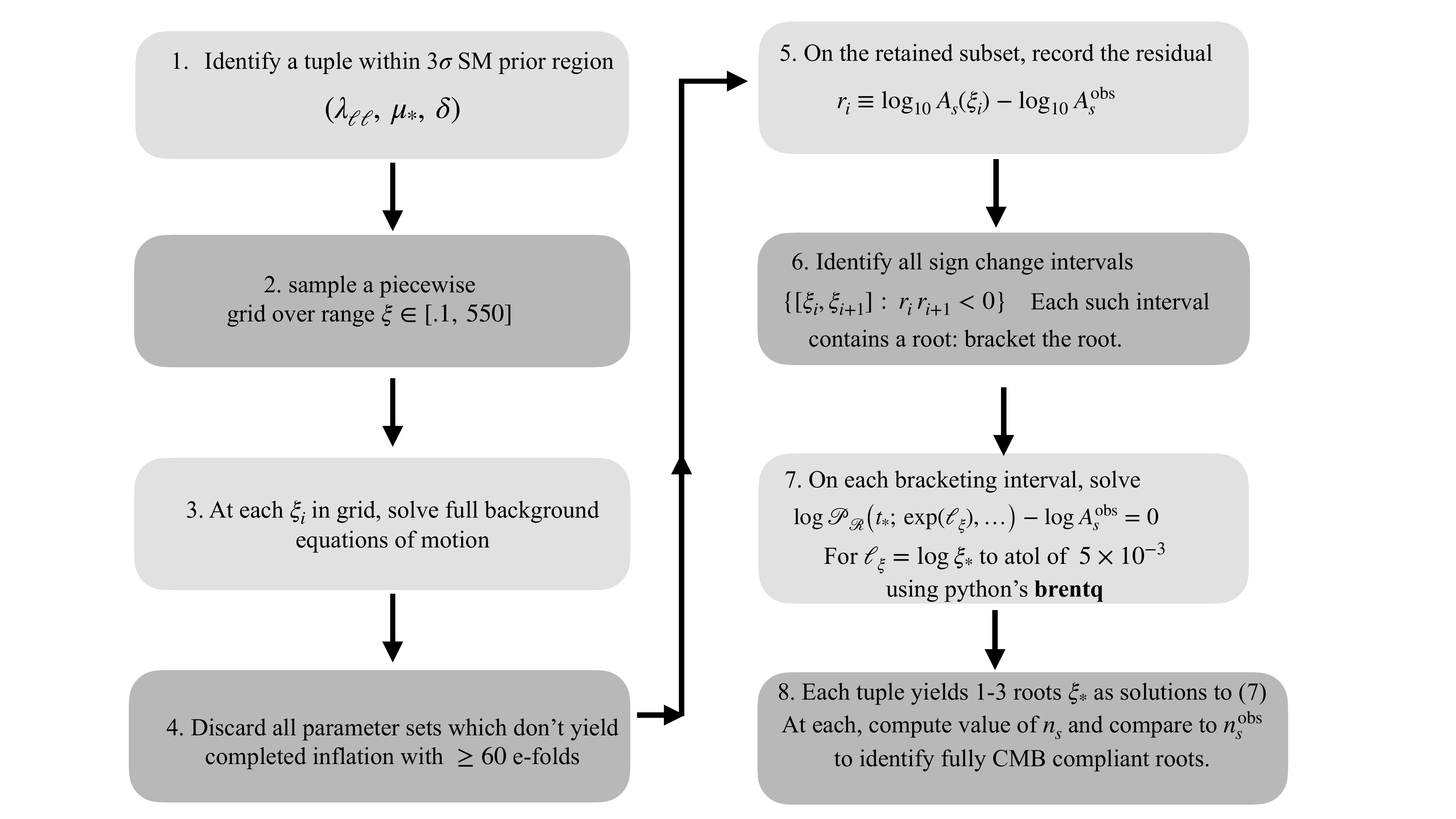}
    \caption{\justifying A visual schematic representation of the algorithm outlined in Appendix~\ref{sec:app_rootfinding} for finding the $\xi_*$ value for a given tuple $\{\lambda_{\rm ll},\mu_{\rm m},\delta\}$ which yields predicted observables within $2\sigma$ of CMB constraints.}
    \label{fig:xiroot_alg}
\end{figure*}
To demonstrate the method for finding the minimum $\xi_*$ to fit the CMB observations given a fixed tuple $\{\lambda_{\rm ll},\mu_{\rm m}, \delta \}$, we will walk through an explanatory example. The reader is encouraged to use the workflow sketched in  Fig.~\ref{fig:xiroot_alg} to follow along at each step. 

Begin by choosing a triplet consistent with the precision EW data. This amounts to a choice of $\{\lambda_{\rm ll}, \mu_{\rm m} \}$ from within the convex hull enclosing the SM $3\sigma$ region shown in Fig.~\ref{fig:chickenpoxplot}. Additionally, choose a small value of $\delta$, generally bounded by $\sim 0.5$, to fix the loop parameter $\lambda_{\rm nl}$. As a worked example, we take the triplet
 \begin{equation}
 \{\lambda_{\rm ll},\mu_{\rm m}, \delta\}= \{2.7\times10^{-5},\,\, 0.0929, \delta_i\}\label{eq:example_params}
 \end{equation} 
where we consider three different values, $\delta_i=(0.01,0.03,0.05)$, to demonstrate that the root-finding method can be applied (in parallel or series) at any value of $\delta$ without loss of generality. 

Within the range $\xi\in [0.1, 550]$, we define a coarse, regime-dependent logarithmic $\xi$-grid consisting of points $\{\xi_{i}\}$. That is, the grid is piecewise-linear, not uniform in log space, with a finer-grained grid at smaller $\xi$. The reason for this adaptive graining will become clear presently. Using our sample point in Eq.~(\ref{eq:example_params}), we define the same initial, static, piecewise grid for each parameter set given in Eq.~(\ref{eq:xigrid_coarse}).
\begin{equation}\label{eq:xigrid_coarse}
      \xi_i^{(0)} \;\in\;
      \begin{cases}      
          \{0.1,\,1.4,\,2.3,\,3.2,\,4.1,\,5.0\}, & \text{6 pts, linear}\\      
          \{5,\,10,\,15,\,\ldots,\,45\},          & \text{step } 5 \\       
          \{50,\,75,\,100,\,\ldots,\,175\},       & \text{step } 25\\        
          \{200,\,250,\,300,\,\ldots,\,550\},     & \text{step } 50 
      \end{cases}                  
\end{equation}  
For each $\xi_i$ in the initial grid, we run the full numerical pipeline to integrate the background field equations of motion and compute all CMB observables. We use the initial conditions described in Sec.~\ref{sec:NumericalResults}, and discard any cases in which inflation never ends ($\epsilon$ never reaches 1).\footnote{While ``never'' is impossible to verify to infinite precision numerically, we set extremely large integration times on the order of 10,000 e-folds as a diagnostic for determining when the field is classically ``stuck,'' either on the plateau or near the inflection point.} Points for which the integrator doesn't return a slow-roll trajectory with at least 60 e-folds of inflation are discarded. 

On the remaining subset of points, we compute and save the residual 
\begin{equation}
     r_{i}\equiv \log_{10}A_s(\xi_i)-\log_{10}(A_s^{\rm obs})
 \end{equation}
at each $\xi_i$, and identify \emph{all} the sign-change intervals $\{[\xi_i,\xi_{i+1}]: r_ir_{i+1}<0\}$. (As noted above, the residual $r_i$ should not be confused with the tensor-to-scalar ratio $r (k_*)$.) By the Intermediate Value Theorem, each such interval must contain at least one root of the equation 
\begin{equation}
     {\cal P}_{\cal R}(t_*; \lambda_{\rm ll},\mu_{\rm m}, \delta,\xi_*)=A_s^{\rm obs} ,
\end{equation}
where, as usual, $t_*$ is time during inflation when modes with $k_*=0.05 { \, \rm Mpc}^{-1}$ exit the Hubble radius. Having obtained an initial set of root candidates, we further refine around each candidate to improve the accuracy of root values and verify that no additional sign changes have been missed by the coarse-graining. For each $\delta$ and each candidate $\xi_*(\delta,j)$, from the initial grid, we add 15 logarithmically-spaced test-values, spanning a factor of two to each side of $\xi_*$, and merge with the original coarse grid into a refined grid.

Detected roots are then sorted and if duplicates are found (including values closer than $1/\xi_*$), the repeated or larger value is discarded. We then reapply the bracketing and root-finding procedure again, just as before, producing the final set of root values. 

Across the 471 SM-consistent grid points of Fig.~\ref{fig:chickenpoxplot}, the residual $\log_{10} A_s(\xi) - \log_{10} A_s^{\rm obs}$ admits either a single root or three roots in $\xi$, because the function $A_s(\xi)$ at fixed $\{\lambda_{\rm ll}, \mu_{\rm m}, \delta \}$ has up to three intersections with the horizontal line at $A_s=A_s^{\rm obs}$, as shown in Fig.~\ref{fig:roots_for_three_deltas}. 

When the sweep produces a single root, the corresponding $\xi_*$ is always fully CMB compliant: $r, \alpha, n_s, A_s$ all lie within the $2\sigma$ Planck/BICEP--Keck bounds. When the sweep produces three roots, the largest is always fully CMB-compliant, the middle is rejected.\footnote{Except in a small number of cases (approximately 0.6\%). In such cases, the root may appear valid but upon closer inspection it is clear that the root should be unphysical: it always occurs when a peak ${\cal P}_{\cal R}(k_{*})$ occurs at CMB scales and is high enough to cross the observed $A_s\sim 2\times10^{-9}$ target value, but is nevertheless accepted because the uncertainty in the observed $n_s$ and tensor-to-scalar ratio are still too large to fully rule it out. This is evident from the bottom-center panel of Fig.~\ref{fig:6panel}.} The smallest root is compliant in some cases and rejected in others.  
\begin{table*}
  \centering
  \begin{ruledtabular}
  \begin{tabular}{l c c c c}
  Category & Strip & $\log_{10}\mu_{\rm m}$ range & $\log_{10}\lambda_{\rm ll}$ range & $N_{\rm pts}/N_{\rm tot}\times100\%$ \\
  \hline
  \multirow{2}{*}{Single root}
      & upper & $[-0.677,\, -0.548]$ & $[-4.728,\, -4.604]$ & 12.31\% \\ 
      & lower & $[-1.483,\, -1.289]$ & $[-4.595,\, -4.480]$ & 12.52\% \\    
  \hline
  Three roots, smallest CMB-compliant
      & --- & $[-1.289,\, -0.935]$ & $[-4.736,\, -4.489]$ & 43.74\% \\  
  \hline
  Three roots, only largest is CMB-compliant
      & --- & $[-0.935,\, -0.709]$ & $[-4.736,\, -4.542]$ & 31.42\% \\     
  \end{tabular}
  \end{ruledtabular}                                                   
  \caption{\justifying Stratification of the 471-point grid bounded by the SM $3\sigma$ region (at $\delta = 0.01$) by root-structure category. The single-root region splits into two disjoint strips at the top and bottom of the blob in $\log(\mu_{\rm m})$; the three-root region forms a contiguous band between them, further subdivided by whether both the smallest and largest or only the largest root passes the CMB-compliance filter on $n_s, \alpha_s$, and $r$.}
  \label{tab:roots}
  \end{table*} 
The parameter space projected onto the $\lambda_{\rm ll},\mu_{\rm m}$-plane  is neatly stratified into bands along $\mu_{\rm m}$ and approximately constant in $\lambda_{\rm ll}$, each with a root-type.\footnote{The apparent dependence on $\lambda_{\rm ll}$ is a result of the irregular shape of the SM $3\sigma$ region, where the extent of $\lambda_{\rm ll}$ depends on the $\mu_{\rm m}$ value.} This structure is summarized in Table~\ref{tab:roots}. 
 
The three regimes of $\mu_{\rm m}\sqrt{\xi}$ illustrated in Fig.~\ref{fig:6panel} have a direct interpretation in terms of the root-finding procedure we have described. For the sample parameter point in Eq.~(\ref{eq:example_params}), the algorithm identifies three roots of $\log_{10} A_s(\xi) = \log_{10} A_s^{\rm obs}$ at $\xi_*= 73.7, 123.2, 206.1$, as shown by the purple interpolating curve and purple stars in Fig.~\ref{fig:roots_for_three_deltas}. 

\section{Self-Consistent Calculation of $N_{\rm CMB}$ from Background Evolution}

In the analysis performed and presented in the bulk of this article, we fixed the value of $N_{\rm CMB}$, the number of e-folds prior to the end of inflation at which the CMB pivot scale first exits the Hubble radius, to be $N_{\rm CMB}=55$. This approximation allows one to run the root-finding algorithm described in Appendix~\ref{sec:app_rootfinding} efficiently for a relatively fine grid over a chosen $\delta$-slice of the $3\sigma$ SM parameter space. 

In principle, however, the pivot scale is identified as the time $t_*$ at which the comoving wavenumber $k_*$ first crossed out of the Hubble radius during inflation, that is, when $k_*=a(t_*)H(t_*)$. We define $N_{\rm CMB}$ as
\begin{equation}
    N_{\rm CMB}= N(t_{\rm end})-N(t_*).
\end{equation}
$N_{\rm CMB}$ is then computed as in Eq.~\eqref{NCMB}, where $V_{\rm E}(t_*)$ is the Einstein-frame potential at horizon crossing, and $H(t_{\rm end})$ is the Hubble rate at the end of inflation, fixed by $\epsilon (t_{\rm end})=1$. $N_{\rm reh}$ absorbs the model-dependent uncertainty in the reheating history; for Higgs inflation, reheating is typically efficient with $N_{\rm reh}\lesssim 3$ e-folds \cite{Bezrukov:2008ut,Garcia-Bellido:2008ycs,Sfakianakis:2018lzf,DeCross:2015uza,DeCross:2016fdz,DeCross:2016cbs,Nguyen:2019kbm,vandeVis:2020qcp,Bettoni:2021zhq,Laverda:2023uqv,Bettoni:2024ixe,Figueroa:2024asq,Chowdhury:2025mye}. In particular, the values of $V_{E}(t_{*})$ and $H(t_{\rm end})$ depend on the value of \emph{all} parameters through the background trajectory, \emph{including} $\xi$. The trajectory, in turn, depends on the choice of the pivot scale: at fixed $\{\lambda_{\rm ll},\mu_{\rm m}, \delta \}$, the value of $\xi_*$ is found (through the finding algorithm presented in Appendix~\ref{sec:app_rootfinding}) by requiring the slow-roll prediction for the scalar power spectrum amplitude, ${\cal P}_{\cal R}(k_*)$, matches the observed value at the assumed pivot $k_* = 0.05 \, {\rm Mpc}^{-1}$,
\begin{equation}
  A_s\left(\xi_* N_*\right) = A_s^{\rm obs}.
  \label{eq:Asroot}
\end{equation}
Eqs.~\eqref{NCMB} and \eqref{eq:Asroot} together constitute a coupled non-linear system in $\{\xi_*, N_*\}$. We solve this problem in two stages.\\
  
\noindent \textbf{Self-consistency}: In the first stage, we perform a simple consistency check.
Self-consistency requires that 
\begin{equation}
  N_{\rm CMB} = N(t_{\rm end})-N(\xi_*) .
  \label{eq:selfconsistency}
\end{equation}
Assuming $N_{\rm CMB}=55$ is fixed, we use the root-finding algorithm detailed in Appendix~\ref{sec:app_rootfinding} to find the CMB-consistent value of $\xi$, and solve the background trajectory with the resulting full set of precision EW-scale measurements and CMB-compliant parameters. We evaluate Eq.~\eqref{NCMB} for this trajectory and compare the computed value of $N_{\rm CMB}$ to the fixed value of $55$. If the resulting value is within a few e-folds, it is also within the uncertainty due to the reheating duration, $N_{\rm reh}\lesssim 3$ e-folds. For metric gravity, the consistency check on 58 evenly distributed points selected from the grid in Fig.~\ref{fig:chickenpoxplot} shows that the computed value is between $N_{\rm CMB}=[56.18,56.98]$ with a mean value of $56.67$, thus the offset is systematic (that is, positive), always slightly underestimating the true value by a few percent, but well within the reheating uncertainty. \\
 
\noindent \textbf{Picard Iteration}: In the second stage, we solve this coupled system directly by Picard (fixed-point) iteration. Given a first estimate $N_*^{(0)}$ of the pivot scale, and parameters $\{\lambda_{\rm ll},\mu_{\rm m},\delta \}$ we use the root-finding algorithm of Appendix \ref{sec:app_rootfinding} to find the value $\xi_*^{(0)}$ for which Eq.~\eqref{eq:Asroot} is satisfied. We then compute a trajectory  from parameters $\{\lambda_{\rm ll},\mu_{\rm m},\delta, \xi_*^{(0)} \}$ and use it to update the pivot by solving Eq.~\eqref{NCMB}: 
\begin{equation}
N_{\rm CMB}^{(1)}= 62
 + \frac{1}{4}\ln\left[\frac{V_E^2(t_*^{(0)})}{3\,M_{\rm Pl}^6\,H^2(t_{\rm
  end}^{(0)})}\right]
  - \frac{N_{\rm reh}}{4} .
  \label{eq:picard-base}
\end{equation}
The updated pivot becomes the estimate for the next iteration. In general, 
\begin{equation}
N_{\rm CMB}^{(k+1)}= 62
 + \frac{1}{4}\ln\left[\frac{V_E^2(t_*^{(k)})}{3\,M_{\rm Pl}^6\,H^2(t_{\rm
  end}^{(k)})}\right]
  - \frac{N_{\rm reh}}{4} .
\label{eq:picard-map}
\end{equation} 
Thus, the first stage is simply Eq.~\eqref{eq:picard-base} of the Picard iteration. We initialize at $N_*^{(0)}=55$ and iterate until convergence, which we define by the condition that
\begin{equation}
       \left|N_*^{(k+1)}-N_*^{(k)}\right|<\varepsilon
\end{equation} 
for fixed tolerance $\varepsilon= 0.05$. The converged pair $\{\xi_*,N_* \}$ then satisfy both Eqs.~\eqref{eq:Asroot} and \eqref{NCMB} simultaneously. 
   
One might be concerned that we cannot be certain the fixed-point iteration will always converge. In fact, convergence to a unique fixed-point is ensured in this instance by the Banach fixed-point theorem. Stated for the case at hand, consider the closed interval $I\subset\mathbb{R}$ of $N_*$ values containing the initial guess $N_*^{(0)}$ and the fixed point $N_*^{(\rm SC)}$, and suppose the Picard iteration map, $T: I \to I$, is a contraction on $I$. A contraction is a map such that there exists a constant $a$ in the unit interval such that 
\begin{equation}\left|T(N_a)-T(N_b)\right|\leq a \left|N_a-N_b\right|
\end{equation} 
for all $N_a,N_b$ in $I$. In our case, it is relatively easy to show that $T$ is a contraction. The Picard-iteration is really a composition of the two steps just described. To see that $T'$ is bounded, we evaluate $T$ across the grid over the SM $3\sigma$ region of Fig.~\ref{fig:chickenpoxplot}, then compute $T'$ by finite differences. We take $d\xi_* / dN_*$, the change in the value of $\xi_*$ identified by the root-finder as we varied $N_*$, then multiply this by $dN_{\rm CMB} /d\xi_*$, the change in the value of $N_{\rm CMB}$ obtained from Eq.~\eqref{NCMB} by plugging in the trajectory for a given $\xi_*$ on the right-hand side. The maximum sampled $|T'|\lesssim 10^{-2}$, showing that $T$ is indeed a contraction. 

By Banach's theorem, $T$ then admits a unique fixed point $N_*^{\rm SC}$ in $I$, that is, a point where $T(N_*^{\rm SC})=N_*^{\rm SC}$. In practice, one cannot check either that $T(I)\subset I$, or that $T$ is a contraction, without knowing where the fixed point is, which is what the iteration is trying to determine. The resolution is to pick the interval $I$ from an \emph{a priori} well-motivated bracket that is wide enough to contain any plausible fixed point, and then verify contractivity. Our \emph{a priori} interval is chosen by the self-consistency checks described as stage 1 in this appendix, which allows us to generously bracket the predicted fixed point. Even without this approximation, standard slow-roll inflation constrains the range of $N_{\rm CMB}$ to lie roughly within $40-70$ e-folds before the end of inflation, marginalizing over reheating histories in order to match the near scale-invariant power spectrum of the CMB. This interval is much wider than the $\mathcal{O}(2)$ e-fold displacement we found between $N_*^{(0)}=55$ and the converged $N_*^{\rm SC}$ in all cases.

Moreover, the Picard sequence generated from \emph{any} initial guess $N_*^{(0)}$ in $I$ will converge to $N_*^{\rm SC}$. In fact, it will converge at the rate of a geometric series, 
\begin{equation}
    \left|N_*^{(k)}-N_*^{\rm SC}\right|\leq \frac{a^k}{1-a} \left|N_*^{(1)}-N_*^{(0)}\right|,
\end{equation} 
where $a$ is the constant defining the contraction $T$, which we find numerically to be $\mathcal{O}(10^{-2})$. Both conditions of the Banach fixed-point theorem are met, which guarantees a unique and self-consistent choice of $N_{\rm CMB}$. \\

\noindent \textbf{Metric gravity}: Comparing to the simple self-consistency check for metric gravity, we find the Picard iteration converges within a maximum of 3 steps to a mean of $N_{\rm CMB}=56.63$, matching the self-consistency method to two decimal places. The final value of $\xi$ consistent with both CMB and SM measurements differed from the naive value at fixed $N_{\rm CMB}=55$ by $1.7\%$ (median), with a spread of only $1.45\%$, while all CMB observables change by less than $0.01\%$. \\

\noindent \textbf{Root-selection discontinuities in multi-root regions in metric gravity}: There is one caveat to the relative stability of the CMB-compliant $\xi_*$ values found by Picard iteration. For some parameter points $\{\lambda_{\rm ll}, \mu_m, \delta \}$ in the SM $3\sigma$ region, the amplitude equation Eq.~\eqref{eq:Asroot} admits multiple roots at fixed $N_{\rm CMB}=55$, as discussed at length in Appendix \ref{sec:app_rootfinding}. Our selection rule picks the smallest root that yields a trajectory satisfying both SM and CMB constraints. In some rare cases, the smallest root will fail $n_s$ compliance at fixed $N_{\rm CMB}=55$, and the algorithm selects the larger root, but a slightly different value $N_{\rm CMB}$ would yield a smallest root that \emph{is} CMB compliant. If the Picard iteration happens to converge to a value of $N_{\rm CMB}$ for which the smallest root becomes viable, the iteration appears to produce a large shift in $\xi_*^{\rm SC}$ relative to the initial guess of $N_{\rm CMB} = 55$. This does not reflect a response in the underlying dynamics; rather, it points to the root-finder legitimately identifying a CMB-compliant root. 

Our analysis suggests this occurs at a rate of less than $2\%$ within the SM $3\sigma$ region, and when it does, the pair $\{\xi_*^{\rm SC},N_*^{\rm SC} \}$ identified by fixed-point iteration remains the correct solution. In such cases, however, the fixed point cannot be treated as a perturbative correction to fixed $N_{\rm CMB}$; instead we identify all such points and treat these separately, which allows us to maintain the contractivity of the mapping on the remaining set of points. \\

\noindent \textbf{Palatini gravity}: We repeat the simple self-consistency check---that is, a single fixed-point iteration step---for Palatini gravity, and find that the resultant $N_{\rm CMB}^{(1)}$ has a mean of $51.87$, and deviates from the fixed value ($N_{\rm CMB}^{(0)}=55$) by $-5.72\%$ (median), with a spread of only $0.48\%$. The resulting shift is $|\Delta\xi_*/\xi_*^{(1)}|\simeq 11.2\%$ (median), with a spread $[10.3\%,11.3\%]$. The shift is thus both systematic and tightly clustered within $\sim 3$e-folds---of order the uncertainty from reheating.  The small shift and tight clustering mean that further iteration would provide only a minimal correction (less than an e-fold). We note, however, that the $\Delta\xi_*/\xi_*$ is about six times larger in Palatini gravity than in metric gravity; this is consistent with the larger shift $\Delta N$ in Palatini and with the scaling of $d\ln \xi_* /d\ln N_*$ which is twice as large in Palatini.\footnote{It is helpful to think about Eq.~\eqref{eq:Asroot} as defining a curve in $\{N_*,\xi_*\}$ space. The slope of this curve determines the response in $N_*$ as one changes $\xi_*$. As a result of the kinetic pre-factors, in metric gravity, the curve scales as $A_s\simeq N_*^2/\xi_*^2$ on the plateau, while in Palatini gravity, it scales $A_s\simeq N_*^2/\xi_*$.}
Once more, $n_s,r,$ and $\alpha_s$ are extremely stable, with median shifts no larger than $10^{-8}$.

\bibliography{ref}

\end{document}